\documentclass[journal]{vgtc}       
\usepackage[svgnames]{xcolor}
\onlineid{0}

\vgtccategory{Research}

\vgtcpapertype{Full paper}

\usepackage{xspace}
\newcommand{\sys}{viz2viz\xspace}

\usepackage{subcaption}
\usepackage{MnSymbol}
\usepackage{enumitem}
\usepackage{url}
\usepackage{pdfpages}

\title{\sys: Prompt-driven stylized visualization generation using a diffusion model}

\author{Jiaqi Wu, John Joon Young Chung, and Eytan Adar}
\authorfooter{
\item
Jiaqi Wu, John John Young Chung, and Eytan Adar are with the University of Michigan. E-mail: {wujiaq,jjyc, eadar}@umich.edu.
}

\shortauthortitle{Wu \MakeLowercase{\textit{et al.}}: \sys: Prompt-driven stylized visualization generation}

\abstract{%
  Creating stylized visualization requires going beyond the limited, abstract, geometric marks produced by most tools. Rather, the designer builds stylized idioms where the marks are both transformed (e.g., photographs of candles instead of bars) and also synthesized into a `scene' that pushes the boundaries of traditional visualizations. To support this, we introduce \sys, a system for transforming visualizations with a textual prompt to a stylized form. The system follows a high-level recipe that leverages various generative methods to produce new visualizations that retain the properties of the original dataset. While the base recipe is consistent across many visualization types, we demonstrate how it can be specifically adapted to the creation of different visualization types (bar charts, area charts, pie charts, and network visualizations). Our approach introduces techniques for using different prompts for different marks (i.e., each bar can be something completely different) while still retaining image ``coherence.'' We conclude with an evaluation of the approach and discussion on extensions and limitations.
}


\teaser{
  \centering
  \includegraphics[width=\linewidth]{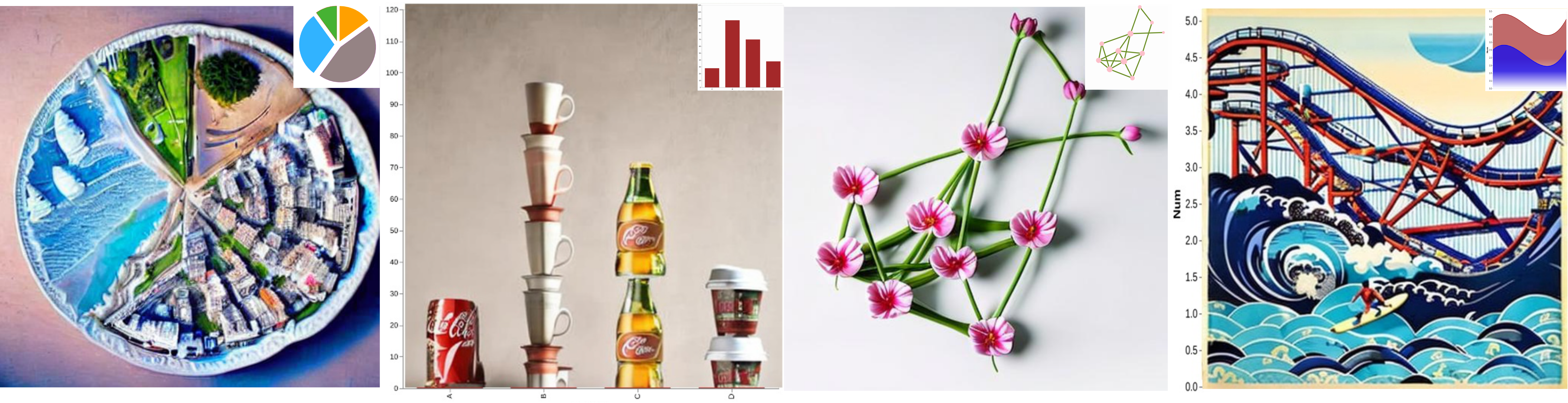}
  \caption{\sys generated samples (original data inset). From left to right: \textit{bird eye view of green forests with many trees, blue ocean with ships, grey city with many buildings, orange deserts}; \textit{realistic stacks of red coca-cola coke cans, brown tea cups, glass wine bottles, starbucks paper coffee cups}; \textit{realistic pink tulips}; \textit{Ukiyo-e style side view of red wooden roller coaster, Ukiyo-e style blue sea waves and surfers}. Note that in some examples in this paper, we select prompts to demonstrate the capabilities of \sys rather than for their aesthetic or design qualities.}
  \label{fig:teaser}
}

\renewcommand{\manuscriptnotetxt}{}

\graphicspath{{figures/}{pictures/}{images/}{./}} %

\usepackage{tabu}                      %
\usepackage{booktabs}                  %
\usepackage{lipsum}                    %
\usepackage{mwe}                       %

\usepackage{balance}
\usepackage{mathptmx}                  %

\begin{document}

\maketitle
\section{Introduction}

Most visualization software aims for the accurate reproduction of data using simple, \textit{abstract} 2-dimensional graphical marks: the rectangles ($\square$) of the bar chart; the arcs ($\leftslice$) of the pie; and the mix of circles ($\medcircle$), rectangles ($\square$), and occasional stars ($\medstar$) in scatter plots~\cite{8807226}.  In many situations, these representations are insufficient for a creative design vision. Graphical `embellishments' require additional styling and design work. A designer might choose to \textit{stylize} a visualization by replacing simple abstract marks with alternative representations: candles instead of bars, slices of fruit instead of arcs in a pie chart, and flowers instead of circles in a scatter plot. Beyond the simple replacement of marks, visualization styling can use alternative artistic forms and media (a sketch, a painting, a photograph, clay, etc.). Our particular focus is on stylized forms that represent a significant stylistic change to how the mark is represented (e.g., stacked coffee cups), but are still recognizable in their original idiomatic forms (e.g., a bar chart)\footnote{While there are many examples of these visualizations, there does not appear to be a unifying name for them (some are called embellished, infographics, photovisualizations). We refer to the broad category as \textit{stylized visualizations}.}. Due to the uniqueness of many of these representations, for a designer to achieve their vision, they need additional tools, skills, and procedures. For example, to manually produce the charts in Figure~\ref{fig:teaser}, a designer would need to create and photograph a physical model or use sophisticated painting tools. Producing a single instance, let alone iterating over a design, is a significantly tedious effort. Instead, we offer \sys, a set of workflows that use generative approaches to produce these images. Our approach allows designers to take ``starting'' visualizations (basic bar charts, area charts, pie charts, and network graphs) and produce stylized visualizations using simple text prompts.

For this work, we build on the systems for generative image creation (e.g., DALL-E2~\cite{dalle2}, Stable Diffusion~\cite{stablediffusion}, and Midjourney). These models are powerful, but on their own difficult to ``contain.'' Without providing any guidance, the output has no relationship to the underlying data. For example, one could ask for ``a photograph of a bar chart made of lit candles with backlighting,'' but the candle heights will not correspond to any actual data distribution. Novel architectures such as img2img~\cite{meng2022sdedit} or ControlNet~\cite{zhang2023adding} allow us to input a starting image (e.g., a standard bar chart). However, these approaches suffer from the opposite problem in that the source image becomes too limiting---the system is unable to produce a stack of coffee cups when guided by a tall 2D rectangle. Our goal is to produce stylized visualizations that are simultaneously flexible to a wide range of creative visions but that at the same time adhere to the underlying distributions of the data. For example, the pie piece that looks like clouds should still be in the correct fraction of the original data. The flowers representing two nodes in a network visualization should be connected if the data calls for it. With \sys, we offer a set of ``recipes'' that can work across a wide array of inputs (both data and prompts) to create both photorealistic and nonphotorealistic images. To identify the range of possible stylized visualizations, we constructed a taxonomy of these forms using human-created examples.

An advantage of our approach is that it does not rely on explicit training using visualization (plain or stylized). This is vital as it is difficult to find a large enough set of examples where we simultaneously have data, descriptions, and stylized forms. Instead, we focus on identifying ways of breaking apart the marks used in the plain form of the visualizations. By `deconstructing' the visualizations as needed, \sys can focus on applying different generative pipelines in a targeted way (e.g., making a bar into the rough shape of a building). Once completed, \sys can reconstruct a cohesive visualization (e.g., multiple buildings side by side) and apply additional generative steps to produce a more composed image (e.g., a more cohesive skyline). In this work, we describe how the basic high-level recipe---composed of sketch, synthesis, and refine steps---can be implemented using combinations of image processing procedures and modified versions of generative pipelines (e.g., depth2Img, img2img, ControlNet, etc.). The \sys approach demonstrates how both traditional image processing (e.g, masking, transformation/translation, etc.) and generative elements can be used in concert to produce a kind of controlled creativity. Beyond the four specific visualization types (bar, area, pie, and network), we offer some guidance on how specific workflows can be built for other types. While our focus in this work is on \textit{textual} prompts as a way to drive the generative system, we describe possible extensions to alternative prompt mechanisms.

Our contributions in this work are:
\begin{itemize}[noitemsep]
    \item Identifying a design space and taxonomy of stylized visualization.
    \item Design and implementation of \sys, a general recipe with specific workflows to support the creation of stylized visualization.
    \item A guide to modifying the \sys `recipe' to different types of charts and prompts.
\end{itemize}

\section{Related Work}

Two lines of research motivate \sys: (1) the visualization literature, with a focus on human- and machine-driven ways of generating stylized visualizations;  and (2) image generation with diffusion models. %

\subsection{Visualization Construction}
Stylized visualizations are primarily created using medium-specific tools \textit{after} being generated through visualization software (e.g., the image file generated by Matplotlib, R, Excel, Altair, etc.). Visualization generators are rarely built to support extensive stylization. For example, in a survey of more than 40 tools, only three supported `infographic customization'~\cite{MEI2018120}).  Instead, designers utilize vector graphics software (e.g., Adobe Illustrator) or `raster' editors (e.g., Adobe Photoshop) to create stylized forms~\cite{10.1145/2598153.2598175}. In response, new tools have been created to bridge visualization software and vector tools (e.g., D3 and Illustrator~\cite{hanpuku}). Inside editors, vector representations can be replaced with icons, vector art, and embellished in various ways. Tools such as Data Illustrator~\cite{10.1145/3173574.3173697}, Charticulator~\cite{charticulator} and (to some extent) Tableau are designed to provide visualization and design feature simultaneously. These allow for the creation of highly stylized vector-art style visualizations. However, none supports bitmap editing that might be required to manipulate photographs or other mediums. With raster images (e.g., photographs), stylization is more complex, but can be achieved by image editors. Image operations (e.g. copy, paste, translating, and transforming) are used to achieve both the desired look, ensuring that the image conforms to the encodings (e.g., dimensions that properly encode the underlying data) and overlaying idiomatic structures (e.g., adding axes or labels). 

An alternative approach to styling visualizations is the application of templates or styles to basic forms. In some cases, the stylized forms are driven by creating hand-crafted examples that are bound to data (e.g.,~\cite{dataink, datadrivenguides}.  For example, DataInk~\cite{dataink} allows end users to draw glyphs by hand (e.g. a sketched tree). Such tools are highly flexible and support different visualization idioms. However, they still require a certain design skill and may involve post-processing work to ensure design variation.  InfoNice~\cite{infonice} represents an alternative approach in which custom (human-designed) glyphs, such as a partially filled wine bottle, are used. Other approaches (e.g., Diatoms~\cite{9552906}) can propose novel but potentially abstract glyph types. Our assumption with \sys is that designers have a conceptualization of the marks they want to generate, which may not be glyphs (abstract or otherwise).

The idea of using machine learning to create visualizations has been applied in different ways~\cite{liu2019latent, wu2021ai4vis, ZHU202024}. However, most approaches are built to suggest visualization types given data or to modify existing visualizations (e.g., annotations or captions). The most related examples offer ways of performing a kind of visualization style transfer (e.g.,~\cite{9585700, 8807266, 9233469, 10.1145/2047196.2047247}). While powerful, these are largely constrained to situations where there are existing examples or where the transformation is from one simple mark type or idiom to another~\cite{7845717}. It is also notable that a consistent limitation in the systems described thus far is that they cannot generate photorealistic visualizations.

The use of photographs or photoimagery in the generation of visualizations is a difficult problem. The DataQuilt system provides one way for designers to work with images to stylize visualizations~\cite{zhang2020dataquilt}. Users can borrow components and stylized features from existing image sources and turn them into elements to which data can be mapped (e.g., resizing an extracted image of a coffee cup based on a data property). Most `found' images may not correspond to the underlying data. For example, a designer would find it challenging or impossible to find or create photographs of objects that are correctly proportioned to the underlying data (e.g., stretched bottles, stacks of cups with correct shadows, etc.).  The Infomages systems proposed an alternative approach where thematic photographs serve as the ``background'' of a visualization (e.g., a picture of lips for a visualization of makeup sales)~\cite{coelho2020infomages}. Infomages allows users to semi-automatically embed traditional chart data into existing images (e.g., coloring different parts of the lips into a pseudo-pie chart). Although the system can find images that may work well with the data, it does not generated these and either `fills' or `overlays' the visualization on top of the image. The interaction between foreground visualization marks and background structure is beyond the scope of \sys.

Our goal for \sys is to start with a standard idiom and use natural language to describe the desired styling. Thus, there is a natural relationship to the use of natural language for visualization construction (see the extensive survey in~\cite{9699035}). Although most tools of this type take language (and data) to generate standard visual forms, some may have a richer set of templates (e.g.,~\cite{8813126}). Systems such as Lumina~\cite{lumina} offer an interesting approach in that they seek to `understand' what the data represents (e.g., a country or specific person) and offer styling based on this understanding (e.g., a country flag or image of the individual). Similarly, MetaGlyph~\cite{ying2022metaglyph} proposes vectorized glyphs based on data semantics (e.g., a layered hamburger if the data is about fast food ingredients). With \sys, we assume that baseline visualization has been constructed in advance (using whatever tool is best) and natural language-based prompts can be used to drive the stylization process without requiring examples or template forms.

Finally, we note that visualization designers have begun to utilize generative approaches to build visualizations. Mostly, this is used for constructing backgrounds or cut-out images that can be placed inside existing marks (e.g., a generated picture of a state's flower for each state in a map). Most experiments note significant additional post-processing steps to ensure that the output conforms to the input distribution. These experiments have not succeeded in mapping to specific idioms or consistently encoding data. However, these experiments have led to incredibly creative works (e.g.,~\cite{JasonForrest2022}) and illustrate a need for better tooling. With \sys, we propose to address this challenge. We define a common `recipe' that supports stylized visualization generation that are simultaneously broad across a wide range of styles and forms, but also ensure consistency to the original data.

\subsection{Diffusion Models}
There are a number of generative image models in use today. However, of particular interest to us are the diffusion probabilistic models that gradually denoise a noise-added variable to get to a sample of interest. This approach was proposed by Sohl-Dickstein et al.~\cite{sohl2015deep} but has evolved significantly in terms of efficiency and steerability. For example, \textbf{Latent Diffusion Model (LDM)}~\cite{rombach2022high} applies the diffusion process over a lower-dimensional compressed representation of images (\textit{latent}) (rather than actual pixel space) for more efficient training. 

With new diffusion models, it is also possible to perform image generation tasks with text inputs, or \textit{prompts}~\cite{dalle2, imagen}. This is achieved by encoding text input into guiding vectors using pre-trained contrasive vision-language models like CLIP~\cite{radford2021learning}. Combining this approach with LDM, \textbf{Stable Diffusion (SD)} is a latent text-to-image diffusion model capable of generating photorealistic images given any text input. As SD is open source, the community has rapidly evolved the approach to generate better results. For example, a parameter that SD users often adjust is \textit{guidance scale}~\cite{ho2022classifier}, which determines how strongly the text input guides the diffusion process.  To allow more controllability, developers devised many \textit{pipelines}, end-to-end image generation processes with an SD model. In constructing \sys, we have adopted a generic recipe to generate stylized recipes. However, depending on the properties of the data, prompts, and desired idioms, each recipe calls for some variations in the diffusion model. In many cases, we leverage and extend existing approaches to fit our specific constraints.

Image-to-Image (\textbf{img2img})~\cite{meng2022sdedit} is a steerable approach conditioned by an additional input image. In img2img, instead of starting the diffusion process from the random latent image, it leverages a noise-added input image as the initial latent. Because of this, the resulting image has visual similarities to the initial image while being influenced by the guiding prompt. In img2img, a \textit{strength} value (0-1) is often used to adjust the level of influence of the initial image.
Based on img2img, \textbf{Inpainting}~\cite{avrahami2022blended} uses an image mask as additional input so that the resulting image is generated only in a specific area.
Other approaches more strictly condition the visual structure of the generated image~\cite{li2023gligen, huang2023composer}.
For example, Depth-to-Image Generation (\textbf{depth2img})~\cite{rombach2022high} allows conditioning of the diffusion process with a depth map of the initial input image (inferred through MiDaS~\cite{ranftl2020towards}). 
Similarly to img2img, depth2img allows users to specify the level of transformation with strength. However, it better maintains the structural information of the original image. 
In addition to these, Zhang and Agrawala~\cite{zhang2023adding} provided a way of adding a conditional control module that accelerates model tuning. 
This approach, or \textbf{ControlNet}, allows various types of condition images (e.g., depth map, edge, or segmentation information) that provide the structural information of generations. For generating stylized visualization, ControlNet with depth or Canny edge detection can be an alternative when depth2img and img2img fail to generate coherent and stylized visualizations.

\begin{figure}[t] 
     \centering

     \begin{subfigure}[b]{0.11\textwidth}
         \centering
         \includegraphics[width=1.1\textwidth]{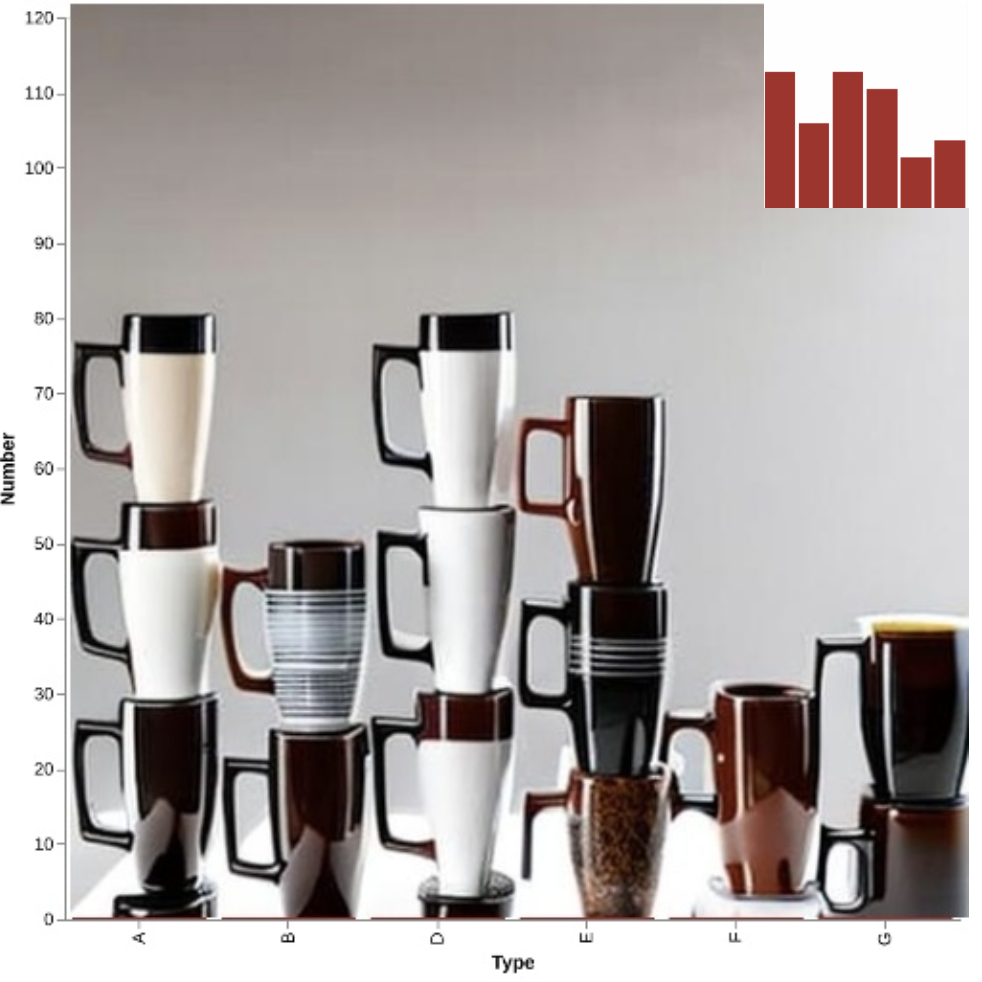}
         \caption{\sys}
         \label{fig:three sin xa}
     \end{subfigure}
     \hfill
     \begin{subfigure}[b]{0.11\textwidth}
         \centering
         \includegraphics[width=1.1\textwidth]{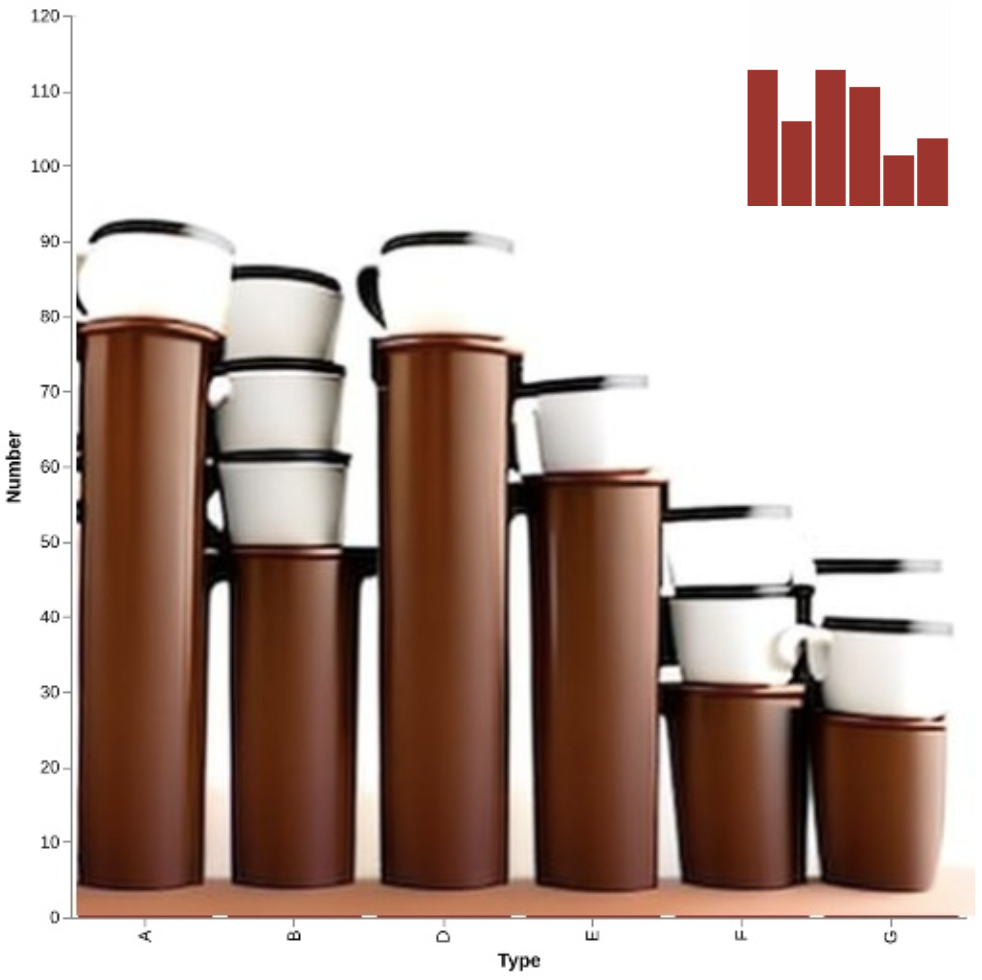}
         \caption{img2img}
         \label{fig:baseline2}
     \end{subfigure}     
     \hfill
     \begin{subfigure}[b]{0.11\textwidth}
         \centering
         \includegraphics[width=1.1\textwidth]{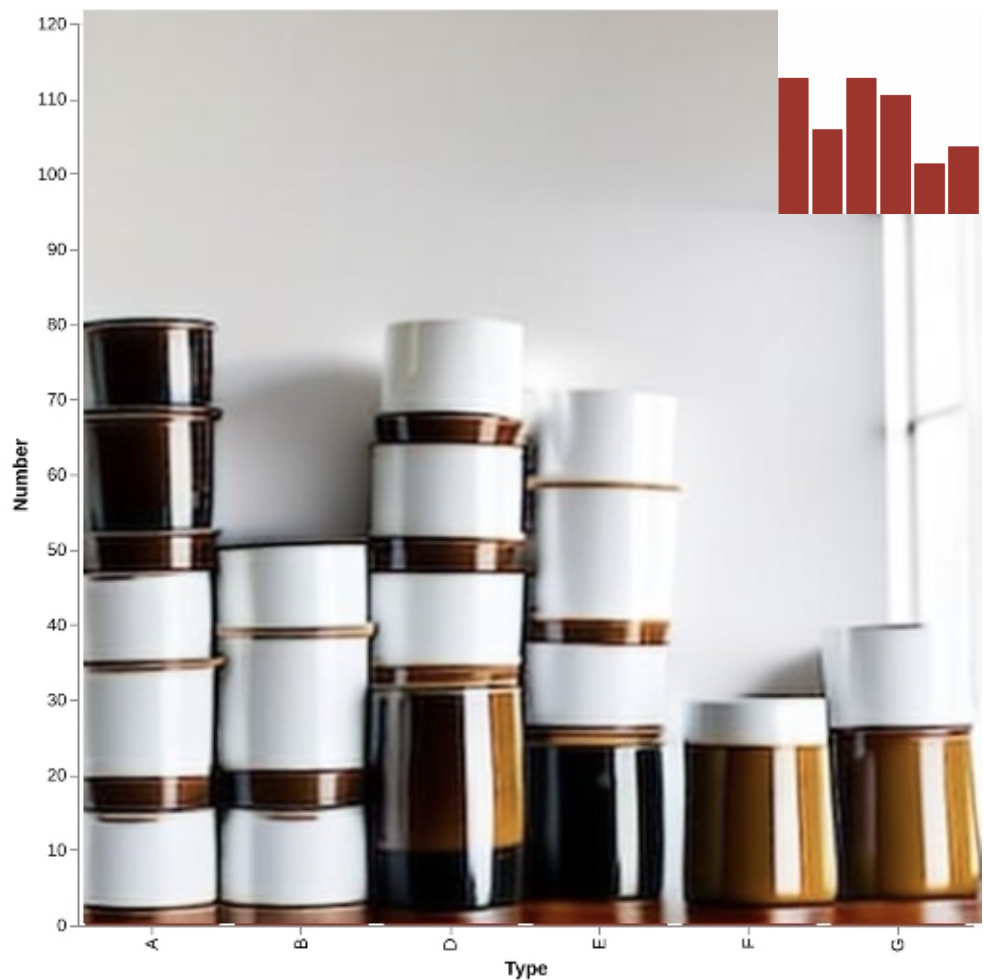}
         \caption{depth2img}
         \label{fig:baseline1}
     \end{subfigure}
    \hfill
    \begin{subfigure}[b]{0.11\textwidth}
         \centering
         \includegraphics[width=1.1\textwidth]{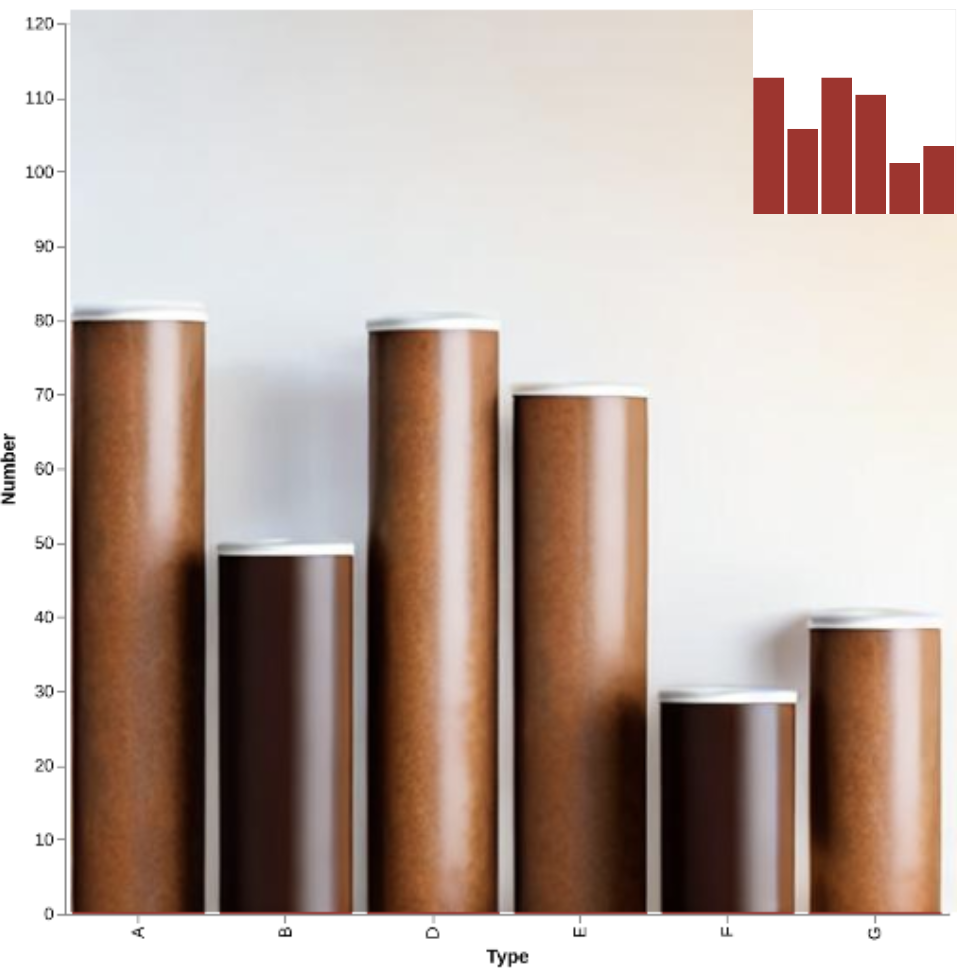}
         \caption{ControlNet}
         \label{fig:y equals xa}
     \end{subfigure}

        \caption{Result for \textit{stacks of realistic coffee mugs filled in with coffee  on a white table} using the different approaches. We compare (a) our system to (b) img2img (strength=0.8), (c) depth2img (strength=0.95), and (d) ControlNet (with Canny edge detection, strength=0.95).}
        \label{fig_baselines}
\end{figure}
 
As we describe in more detail below, applying existing models directly to the visualization problem is problematic. It is difficult to both meet the stylization prompt and preserve the precise information of the chart. Applying these models alone does not produce reliable results (Figure~\ref{fig_baselines}).
For example, with low strength, img2img (Figure~\ref{fig_baselines}b) keeps the structure information of the initial image (a good thing), but is unable to use the prompt to transform the bars (it simple adds``noise''). With high-strength, the image may be transformed but will not adhere to the underlying data distribution. 
Structure-conditioned approaches maintain visualized data, but often follow structure information too strictly and will not stylize the shape of the marks (e.g., Figure~\ref{fig_baselines}d). 
Moreover, if the user wants a visualization with multiple objects as marks with specific sizes, positions, and shapes (e.g., the first bar should be coffee cups, the second should be bottles, etc.), existing controllable approaches would struggle to meet these requirements. With \sys, we can generate stylized visualizations more reliably by combining and modifying these existing pipelines. Our output adheres to both the data and the prompt.

\section{Taxonomy of Stylized Visualization}
\begin{figure*}[ht!] 
     \centering\includegraphics[width=1\linewidth]{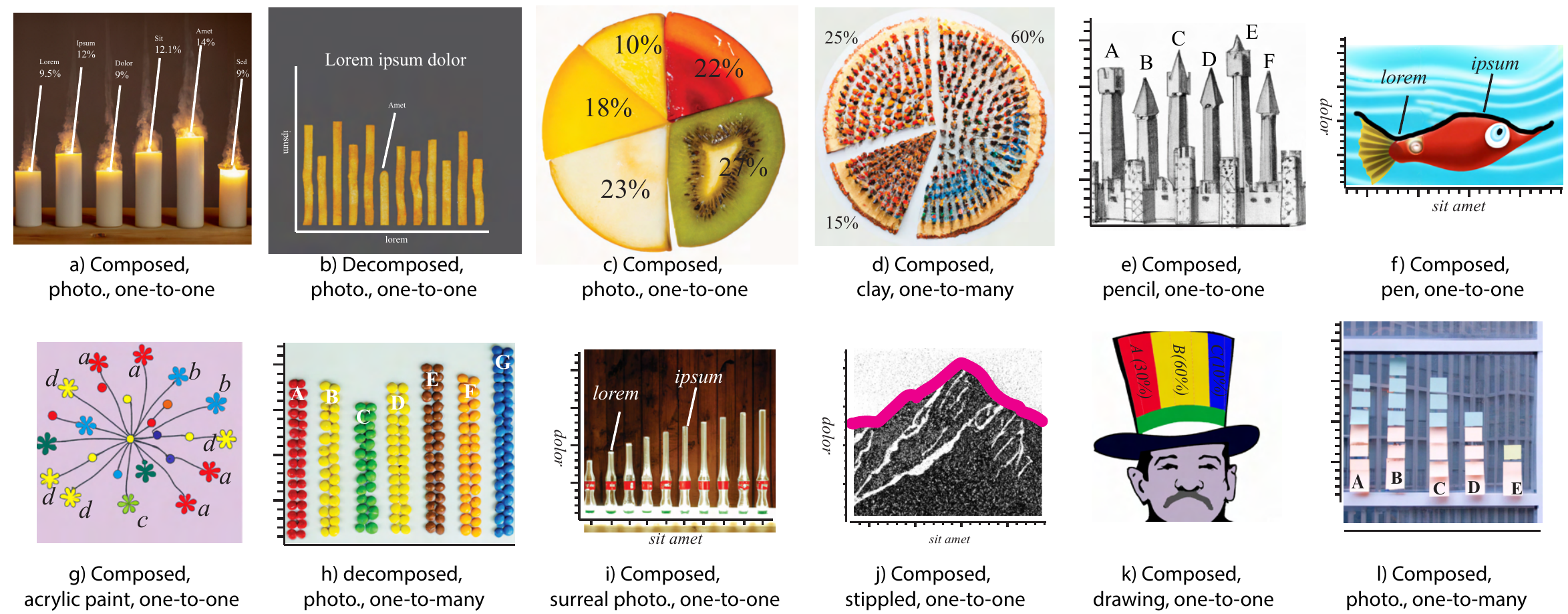}
    \caption{Various examples of stylized visualizations based on a wide set of `found' categories. These were created by us using DALL-E 2 (e.g., \textit{A photograph of a bar chart made of lit candles on a table with backlighting}) but without specific data prompt. Generated images were modified in Illustrator and Photoshop for annotation (e.g., labels, axes).}
    \label{fig:taxonexamples}
\end{figure*}

To understand the space of stylized visualizations we collected over 250 examples. We included those where standard idioms were used but where marks were `concrete.' That is, objects other than abstract geometric structures were used to represent the data. To collect these, we searched for terms such as `infographic' and `photoviz' and included those that matched our definition. We used Pinterest's recommendation engine to find additional examples. 

The authors grouped these images and found three main categories. Figure~\ref{fig:taxonexamples} illustrates a number of synthetic variants based on real-world examples. These were created using DALL-E 2 with some modification in Illustrator to add labels and emphasize certain marks. The images demonstrate the potential of generative approaches. However, as we did not specify the data for these examples, the distributions are meaningless. We compare these examples to those generated by \sys in the supplement.

It is worth mentioning what we excluded. Some visualizations may have been stylized, but under a broader definition than we use here. Work by Federica Fragapane, Giorgia Lupi and Stefanie Posavec can be considered highly stylized. However, their designs often leverage novel abstract forms or idioms. We also do not consider abstract marks translated to the same mark in another medium (e.g., a bar chart painted in oil paints).  These are examples of `style transfer' rather than the stylization we are interested in (for these, a simple use of ControlNet might suffice).  Below, we describe the three main labels we apply to our examples.

\subsection{Style and Medium}
Style and medium refer to the artistic style used and the medium (paint, photography, sculpture, etc.). Due to their extreme variations, we do not create sub-classes within this category (classifying art is well outside the scope of this work). However, we have found recurrent themes in the found images. 

Many of the designs we identified used photographs of physical objects for visualizations. Artists such as Sarah Illenberger,  Peter {\O}rntof, and Oliver Uberti, and the Data Made of Things project (\url{https://datamadeofthings.tumblr.com/}) are representative of this style of design. In Figure~\ref{fig:taxonexamples} examples include images \textit{a-c, h, i}, and \textit{l}. We view this type as distinct from images made using a different medium (e.g., clay) which is then photographed.

Non-photographic stylization often involves some sort of paint or drawing medium~\cite{sturdee2022data}. In real-world examples, these visualizations are usually hand-drawn (digitally or on physical medium). The data journalist Mona Chalabi produces examples of this type, and some of Nigel Holmes' work can also be considered in this category. Examples in the figure include \textit{e, f, g, j} and \textit{k}. The attractiveness of this type is that they allow for the creation of non-photorealistic representations (e.g., a time-series as the back of a drawn fish). While it is possible to conceive of surreal objects as photographs (e.g., the stretched bottles in \textit{i}), designers and viewers may prefer a non-photographic medium.

While we note a few common themes in hand-crafted stylized visualizations, we emphasize that there are constraints that may have led to their popularity. For example, certain cultural aesthetic preferences and norms may have resulted in an over-representation of certain image styles. More critically, certain images may simply be too difficult to create without a generative tool. For example, a visualization that looks like a photograph but is surrealist, may not be possible without extensive image editing work. Because they are challenging to create, we may simply not see many of them. Generative approaches may change the landscape of stylized visualizations.

\subsection{Composed and Decomposed}
\label{composed and decomposed}
Our second category is based on \textit{scene cohesiveness}. Specifically, we are interested in whether the visualization is \textit{composed} or \textit{decomposed}. With composed images, objects representing the marks and backgrounds interact in some way (i.e., objects sit on top of each other, shadows from one bar chart element fall onto another, etc.). Examples of this category include Figure~\ref{fig:taxonexamples} \textit{a, c-g, i} and \textit{j-l}. In these, objects either ``touch'' or are part of the same scene (candles on the same table, post-it notes in the same window, or bars in the same hat). 
Decomposed images may have similar marks but the objects do not interact with each other or their background. For example, if we take separate photographs of each element in the bar chart and cut and paste them into one image, we would consider this a decomposed visualization (e.g., Figure~\ref{fig:taxonexamples} \textit{b} and \textit{h}).
 
\subsection{One-to-One and One-to-Many}
\label{one-to-one-one-to-many}
Finally, we categorize visualizations with how a mark is constructed. Specifically, we are interested `of what things' it is made. In many situations, one mark in the original idiom is represented by one mark in the stylized visualization. One bar to one candle or one French fry or one mountain to one time-series are examples (see Figure~\ref{fig:taxonexamples} \textit{a-c, e-g} and \textit{i, j, k}). 
Alternatively, one-to-many examples are those in which one mark is composed of smaller `parts' such as one pie slice as a crowd or one bar chart as many candies (see \textit{d, h} and \textit{l}). one-to-many examples are often seen as pictograms where icons are used to construct a `mark.'

In some instances, we may find ambiguity within this category. For example, we may use a mountain range to represent a time-series. The range (one) is composed of multiple mountains (many). In these situations we opt to classify the visualization based on what we viewed as the `simplest' (in the gestalt sense) visual explanation (in this case, the whole range).

Our goal for \sys is to support the generation of as broad a range of these stylized visualizations as possible. In addition to ensuring that the marks encode the data correctly, we would like to generate both constructed and deconstructed images, photo-realistic and non-photorealistic, and marks made of both single and multiple objects.

\section{\sys: Generating Stylized Visualization}
\begin{figure*}[htbp] 
     \centering\includegraphics[width=1\linewidth]{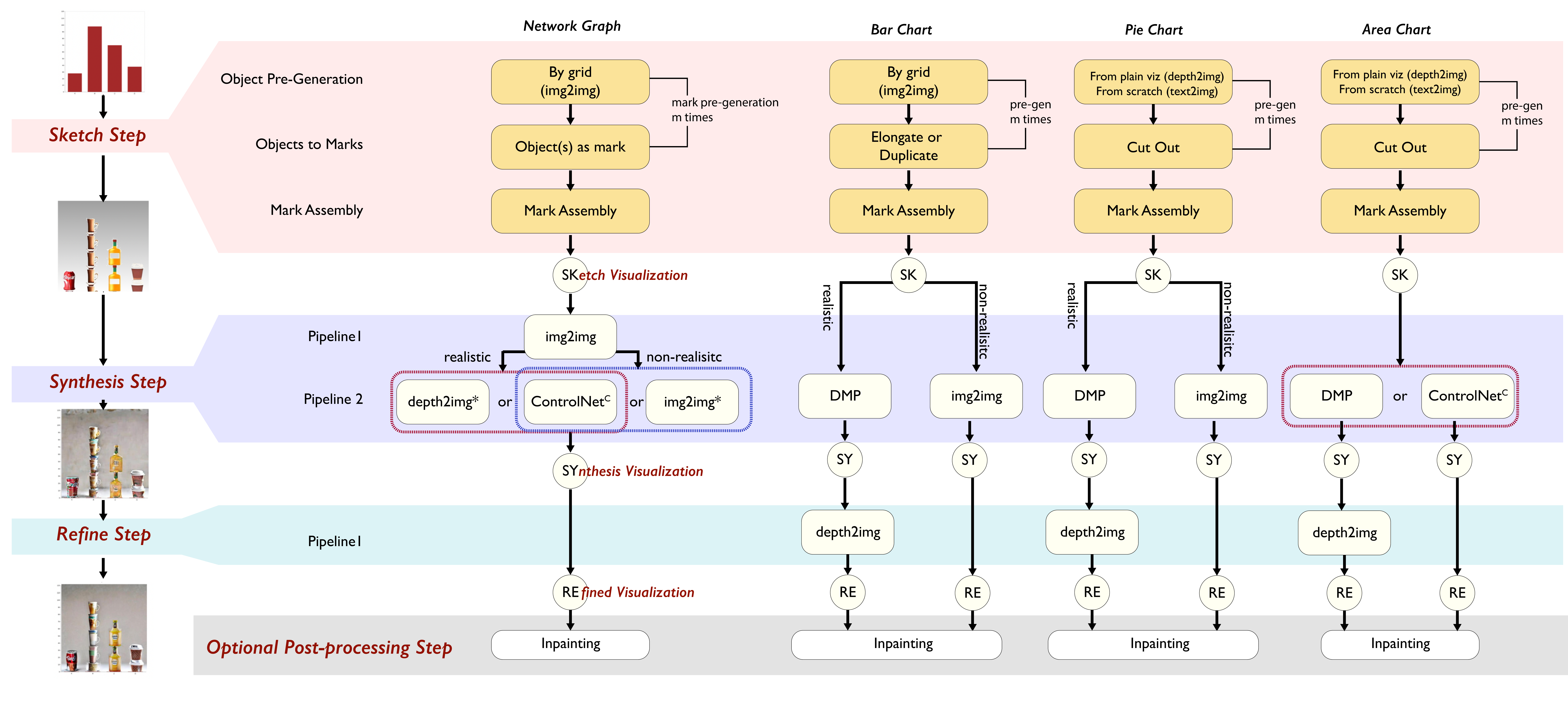}
    \caption{Workflow of \sys. The leftmost path describes the generic recipe, with specific variants appearing to the right. Note that ControlNet\textsuperscript{C} means the ControlNet pipeline with Canny edge condition, and depth2img*/img2img* mean the Edge-enhanced-depth2img/img2img pipelines, which add an image of the network edges on top of the the initial image before the depth2img/img2img pipelines.}
    \label{fig_wholediagram}
\end{figure*}

We introduce \sys, a workflow that generates stylized visualizations using diffusion-based image generation models. Due to the wide variation of visualization inputs and stylized outputs, the core of \sys is a central ``recipe'' that can be adapted to different scenarios. 
\sys is based on the same 3-step framework, which combines existing and novel architectures and pipelines (Figure \ref{fig_wholediagram}). 
While there is a central core recipe, we find that no single workflow works for all visualization types and all prompts. Instead, we offer a number of alternative pathways that the designer can use to achieve their desired result. We leave \textit{completely automated} selection of a workflow's sub-routines for future work. Once defined, a \sys workflow can be run end-to-end without intervention. However, designers can intervene at different places to achieve different effects (e.g., modifying a visualization sketch before running the rest of the workflow).

As input, \sys receives the plain visualization, prompts, and additional control variables. The plain input visualizations (bar chart, pie charts, etc.) can be generated through various tools (in our implementation we use a variety of Python-based toolkits). Prompts are textual descriptors of the stylization we would like to apply. In \sys, prompts are composed of different parts that can be applied selectively in the diffusion steps. Specifically, a prompt consists of: (a) contextual prompt that will apply to the entire process (e.g., \textit{a photograph of a bar chart}), (b) one or more sub-prompts describing what the marks should be transformed to (e.g., all bars should be \textit{lit candles} or bar one should be a \textit{skyscraper} and bar two should be a \textit{pagoda}), (c) an optional background prompt (e.g., \textit{a library} or \textit{a simple gradient}).

\subsection{The \sys Recipe}

The first step of \sys is \textbf{Sketch}. In this step, \sys creates a roughly stylized visualization or sketch. 
As we describe below, \sys breaks each mark (and background) into individual components and generates a stylized representation for each (e.g., each bar in the bar chart is transformed with the prompt). It ensures that marks do not get blended together, which can happen if they are close together or touching. Once re-assembled, the sketch visualization may not be visually coherent as each part may have different lighting or variations in style. The sketch visualization is then passed to a second step: \textbf{Synthesize}. Here, the visualization is made coherent by using the diffusion pipelines to the entire image. Figure~\ref{fig_wholediagram} (left) shows the general workflow with screenshots of what the visualization looks like at each step.

Although visualization after the synthesis step tends to be visually more coherent than sketch visualization, we have found that in some cases an additional \textbf{Refine} step is necessary to clean up the image (e.g., blurriness). This final pipeline produces better image resolutions and stronger details.  Unlike synthesize, the pipelines minimally transform the image to improve quality while preserving visualized information. Below, we describe each step in detail.

\subsubsection{The Sketch Step}

The sketch step includes: 1) \textbf{Mark Pre-Generation}, which transforms each mark in the plain visualization into a stylized object; 2) \textbf{Background Pre-Generation}, which generates a stylized background; and 3) \textbf{Mark Assembly}, which assembles pre-generated marks and background into a sketch visualization. 

\paragraph{Mark Pre-Generation}

In \sys, we use several approaches to generate each mark. As discussed in Section~\ref{one-to-one-one-to-many}, a `single' mark can actually be different things: a single object (a single bottle representing the entire mark), multiple objects (multiple stacked cups), 
or a cohesive part of a larger object (e.g., a pie chart). We describe different approaches to this problem below.

\begin{figure}[t]

\includegraphics[width=\linewidth]{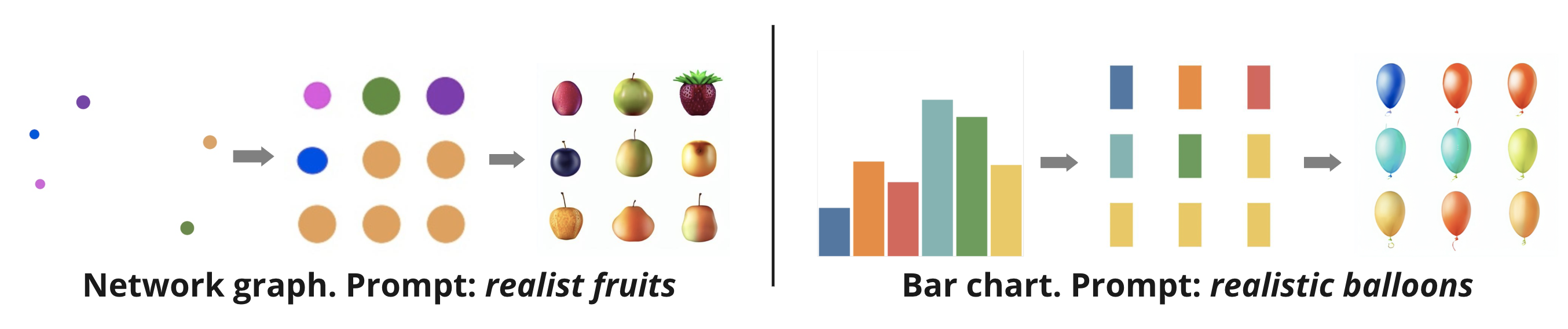} 

\caption{Pre-generation by grid in bar chart and network graphs.}
\label{fig_grid}
\end{figure}

\begin{figure}[t]
  \centering
  \includegraphics[width=\linewidth]{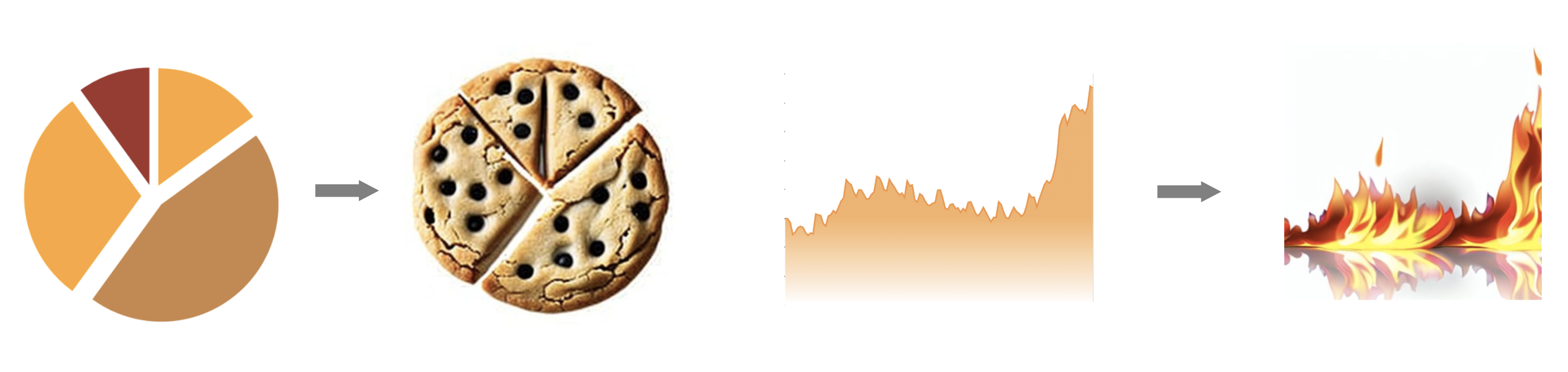} 

\caption{Direct pre-generation for pie charts (prompt: \textit{realistic slice of cookies}) and area chart (prompt: \textit{fire flames}).}
\label{fig_plain}
\end{figure}

\textbf{Pre-generating Objects:} In many cases, when `sketching' replacement marks, \sys begins by utilizing the marks created in the original plain visualization: bars and pie slices are separated, and nodes and edges are split. Depending on the number of marks (e.g., many bars or many nodes), \sys will position the marks in $N \times N$ grids. By creating a clear separation, we can ensure that individual objects are created for each original mark. For example, creating a \textit{social network graph made of fruits} requires many fruits, each somewhat aligned with the original color and size of the node. Each original node mark---for example, a circle---from the plain visualization is placed in a $N \times N$ grid. With this modified image, we can perform an img2img diffusion (Figure~\ref{fig_grid}a). We have found through experimentation that in many situations, it is not necessary to scale the separated mark to the correct relative dimensions. For example, with a bar chart, we simply need the right number of bars, but they can be equal-sized (Figure~\ref{fig_grid}b). The correct scaling can be recovered in later steps. Although the grid approach is effective in many situations, we have found that it is not vital in certain situations. For example, we do not need to build a grid when the number of marks is limited (e.g., a single dataset area chart), where there is sufficient separation in the objects (e.g., pie slices that are not touching), or when we want the generated image to very strictly follow the contours of the original mark (e.g., an area chart). In these cases, we can use the original marks, in their original layout, as input to the pre-generation diffusion phase (see Figure~\ref{fig_plain}). We note that as \sys conditions the generation with the mark, image-conditioned diffusion approaches (e.g., img2img or depth2img) are the most appropriate for this step.  For situations where the prompt consists of multiple shape sub-prompts (e.g., we want different objects for each bar) the sketch pre-generation step is run multiple times, once for each sub-prompt.

In some cases, pre-generation using the original mark may not produce a good sketch. For example, this occurs when the prompt asks for an unconventional camera angle or a mark composed of many items (e.g., \textit{a pie chart slice made of a birds-eye view of many individuals in a crowd}). In these cases, \sys can produce a better sketch by not being restricted to the original mark. However, as this generated image is not constrained by the original mark's shape we need to transform it.

\begin{figure}[t]
  \centering
  \includegraphics[width=0.47\textwidth]{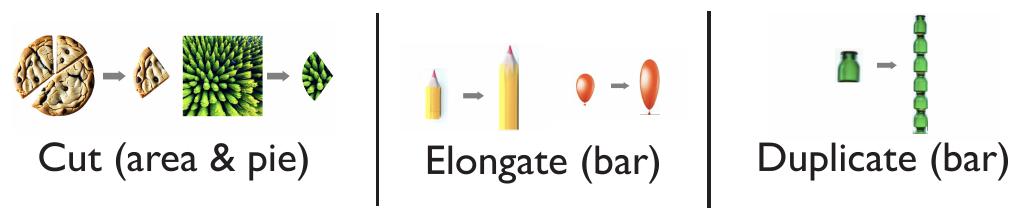} 

\caption{Image operations to translate object(s) to mark(s).}
\label{fig_shapemodify}
\end{figure}

\textbf{Object(s) to Mark(s):}
\sys turns generated objects into one or multiple stylized marks in a number of ways: a) turning an object to a mark directly (e.g., a generated flower $\rightarrow$ a network graph node); b) adjusting multiple generated objects (e.g., a stack of coke cans) so that they create a mark (a stack of coke cans $\rightarrow$ a single bar in a bar chart); or c) modifying the generated object to fit within the constraint of the original mark (e.g., an apple pie $\rightarrow$ a slice of pie). These are all done through various combinations of image transformations (see Figure~\ref{fig_shapemodify}).

In the simplest case, no transformation (beyond translation and basic rescaling) is needed. For example, if we want each node of a network graph to be stylized as an apple, we can simply generate a set of apples (by original mark grid) and then replace the nodes with these images. 

When only parts of the generated objects are useful as marks, \sys will simply cut out a piece of the object by using a mask on the generated image (see cut-out in Figure~\ref{fig_shapemodify}). We found this approach most useful for more complex shapes (e.g., pie slices or time series in an area plot). This approach allows us, for example, to generate an entire apple pie or a large picture of a crowd, and cut out piece corresponding to the pie chart's mark.

In some cases, the produced mark is not in the correct dimensions and cannot be rescaled easily (i.e., both width and height consistently rescaled). When a mark's dimensions require the generate image to have an alternative dimension, \sys performs elongation (Figure~\ref{fig_shapemodify}). Note that in many situations we do not want to stretch the entire object. For example, we might want to only elongate a specific part of the object (e.g., the body part of a pencil, not the tip or eraser). In our current implementation, \sys does not understand the semantics of a generated object. Instead, we adopt a heuristic that works well in practice: the generated object is split into three equal parts (the `head', `body' and `tail'). When applying this workflow, \sys will only elongate the body (central third) of the generated object to the needed dimension while keeping the head and tail dimensions fixed.

When the design calls for `stacking' objects (e.g., stack of coffee cups), \sys can duplicate generated objects and place them on top of each other (see duplication in Figure~\ref{fig_shapemodify}). Pie chart stylization can also use this approach to fill the pie area with duplicates of a specific object.

\paragraph{Background Pre-Generation}
If a background prompt is provided, \sys uses txt2img to generate a background for the sketch visualization. Because backgrounds can visually conflict with the, arguably more important, visualization marks, \sys blurs and brightens the background. If no background prompt is provided, \sys can generate a simple gradient background in this step.

\paragraph{Mark Assembly}
After creating a sketch for each mark, \sys reassembles these marks by placing each corresponding generated image into the position specified by the orignal visualization. For example, \sys will `disassemble' the grid of generated objects and reassemble them into the appropriate locations in the network graph.

\subsubsection{The Synthesize Step}
The sketch produced by the earlier step of \sys generally contains the marks in generally the right shapes, sizes, and placement. However, these images are not cohesive: cropped parts of images may feel incomplete, cups may be floating in space, marks may have lighting and style variations that do not make for a good design. A Synthesize step is used to remedy many of these issues. As with the Sketch step, there are a number of alternative workflows that can be used depending on the input visualization type and prompt. Because of the difficulty in maintaining coherence among the connected elements of a network (which are not always captured in the sketch step), those visualizations require a two-step process. For network graphs, the Synthesize step first uses an img2img pipeline with low strength level. This smooths unnatural object renditions and compositions visible in sketch visualization. 

All visualizations at this point undergo the same (conceptually) Synthesize step to ensure that the various marks can look more coherent. In reality, different pipelines can be used here to achieve different effects. We return to why someone might prefer one pipeline over another when discussing specific chart types. In addition to img2img, depth2img, and ControlNet pipelines that can be applied to the sketched visualization, we also designed three additional pipelines: \textbf{Diffusion with Multiple Prompts (DMP)}, \textbf{Edge-enhanced depth2img} and \textbf{Edge-enhanced img2img}. The first, DMP, is used to support designers who want different objects for each mark (i.e., with multiple sub-prompts). The latter two are used to ensure that edges are retained in network visualizations,  as thin edges can easily vanish during a diffusion process. 

\paragraph{Diffusion with Multiple Prompts (DMP)}

\begin{figure}[t] 
     \centering\includegraphics[width=\linewidth]{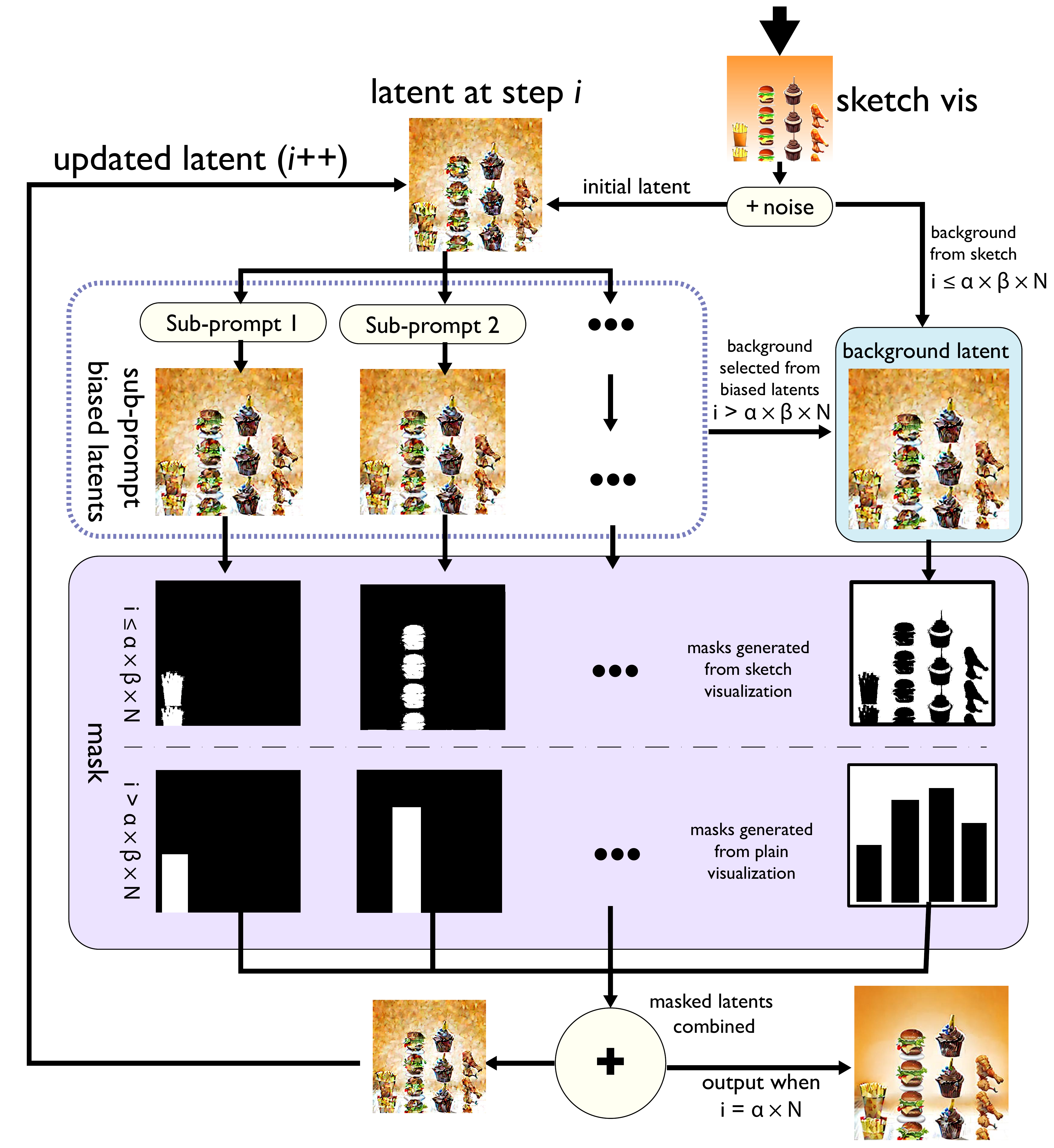}
    \caption{Example illustration of the DMP pipeline. In this example, there are four sub-prompts:  \textit{fries}, \textit{hamburgers}, \textit{cupcakes}, \textit{fried chicken}}.  
    \label{fig_DMP}
\end{figure}

The DMP pipeline is used when we would like to specifically apply different sub-prompts to different marks. With standard pipelines, multiple prompts may lead to blending that is not desirable. For example, if we synthesize a pie chart with the prompt \textit{chocolate and apple pie slices} with img2img or depth2img, we may find slices that combine both apple and chocolate together. Approaches like inpainting and masked application of diffusion to a region of the image will not work here as each area will largely ignore its neighbors and the image as a whole. This is very much against the goals of the Synthesize step where we would like a coherent image--just one made of different objects in this case.

Our solution is a \textbf{Diffusion with Multiple Prompts (DMP)} pipeline. DMP can link different marks in the visualization to different sub-prompts in a single diffusion process, while minimizing the mix of different prompts in a single mark (Figure~\ref{fig_DMP}). The basic idea of DMP is to perform diffusion on the whole latent separately for each prompt and then combine the multiple resulting latents back into a single latent. Since this process is done iteratively for each step, the different areas guided by different sub-prompts could also consider the rest of area, resulting in a coherently synthesized visualization. 

At the start of the DMP pipeline, the process is similar to img2img. DMP adds a specific level of noise (according to the strength $\alpha$) to the sketch visualization latent. This becomes the initial latent. 
For each step (i = $ 1 \cdots \alpha \times N$, where $N$ is the number of whole diffusion steps when generating an image from a pure noise),
DMP diffuses each mark separately using depth2img with different sub-prompts (e.g., `hamburger', `fries', etc.). The result is a set of latents, one per mark. In our example, each of these latents becomes biased towards one of the sub-prompts. This ensures that marks begin to `synthesize' together (i.e., the stack of hamburgers begins to influence the stack of fries) but also means they blend (i.e., the hamburger stack becomes more fries-like). To prevent the latter problem, DMP masks each bar. 
With $\beta$, which is a hyperparameter between 0 and 1, when $i \le \alpha \times \beta \times N$, the image is filtered using a mask generated from the sketch visualization (ensuring a tight adherence to the shape of the sketch). When $i > \alpha \times \beta \times N$, we switch to a mask defined by the plain visualization. This larger mask allows for slightly more flexibility as the diffusion has stabilized. Similarly to each mark, we apply the same idea to the latent for the background. When $i \le \alpha \times \beta \times N$, the background latent is the sketch visualization latent with added noise (and the mask is derived from the sketch visualization).  When $i > \alpha \times \beta \times N$, the background latent is selected from one of the prompt-biased mark latents. We choose the background latent iteratively (switching between sub-prompt outputs) at each step to minimize the influence of any one sub-prompt on the background. Intuitively, this allows the background to develop independently based on the sketch in early phases and then becomes more directly impacted by the newly diffused marks. The masked latents from the marks and background are combined, and this becomes the latent for step $i + 1$. This strategy has two benefits: (1) it reduces reduces computational complexity (i.e., no additional diffusion for the background); and (2) we get higher visual coherence than diffusing the background with a separate prompt. Note that if there is only one sub-prompt (e.g., all bars have the same prompt), DMP reduces to depth2img.

\begin{figure}[t] 
     \centering
     \begin{subfigure}[b]{0.15\textwidth}
         \centering
         \includegraphics[width=1\textwidth]{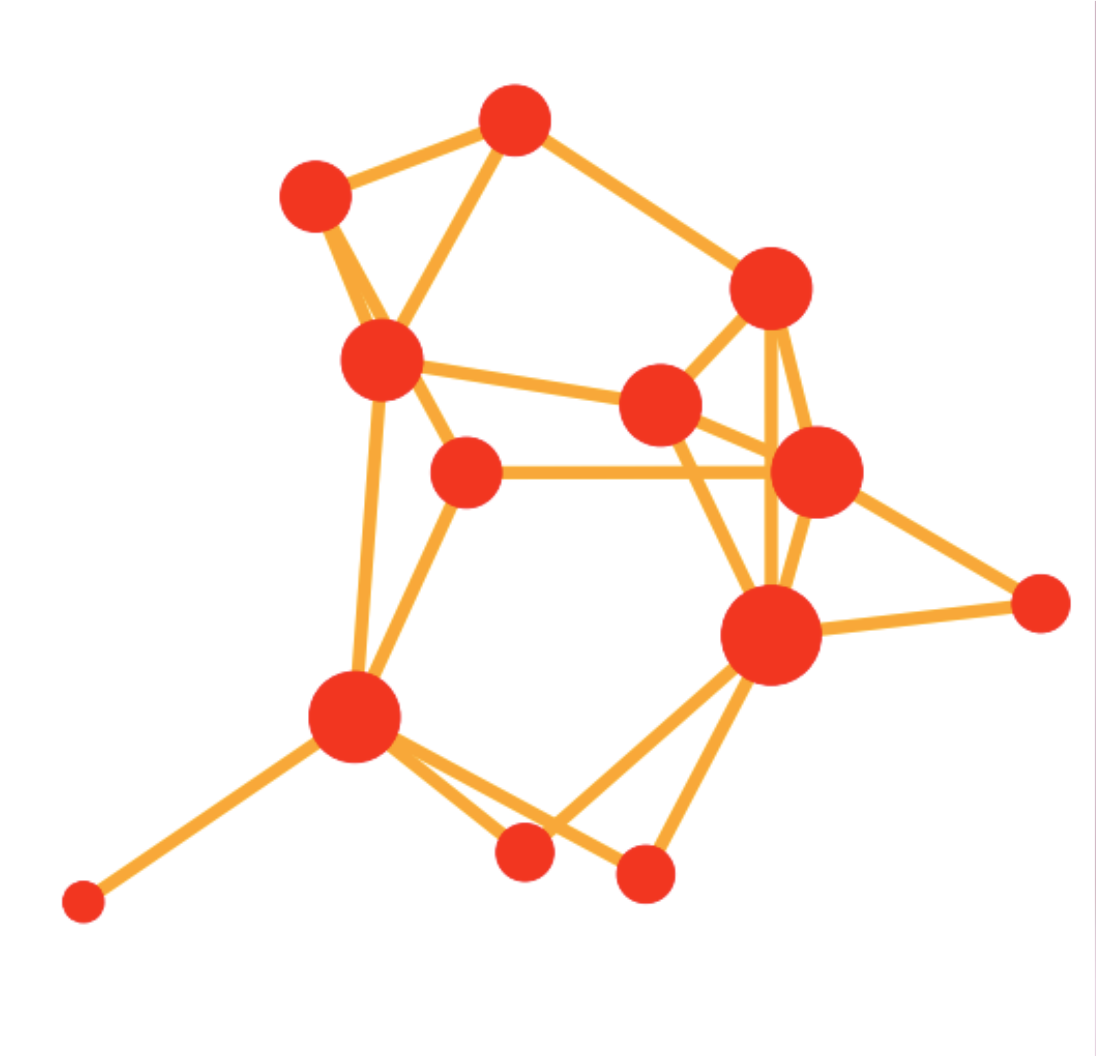}
         \caption{plain visualization}
         \label{fig:y equals x9}
     \end{subfigure}
     \hfill
     \begin{subfigure}[b]{0.15\textwidth}
         \centering
         \includegraphics[width=1\textwidth]{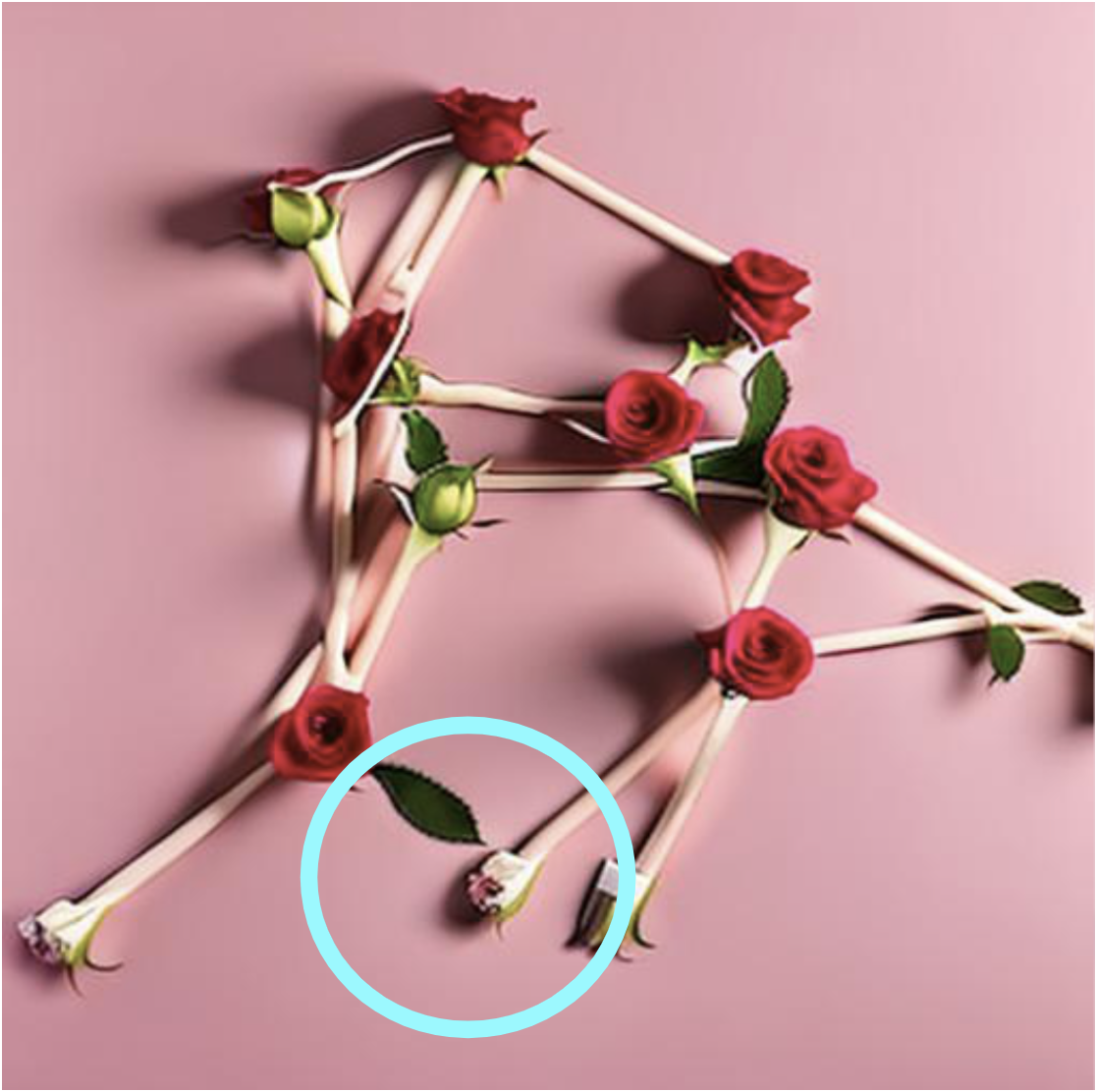}
         \caption{\sys result using Depth2Img}
         \label{fig:three sin x9}
     \end{subfigure}
     \hfill
     \begin{subfigure}[b]{0.15\textwidth}
         \centering
         \includegraphics[width=1\textwidth]{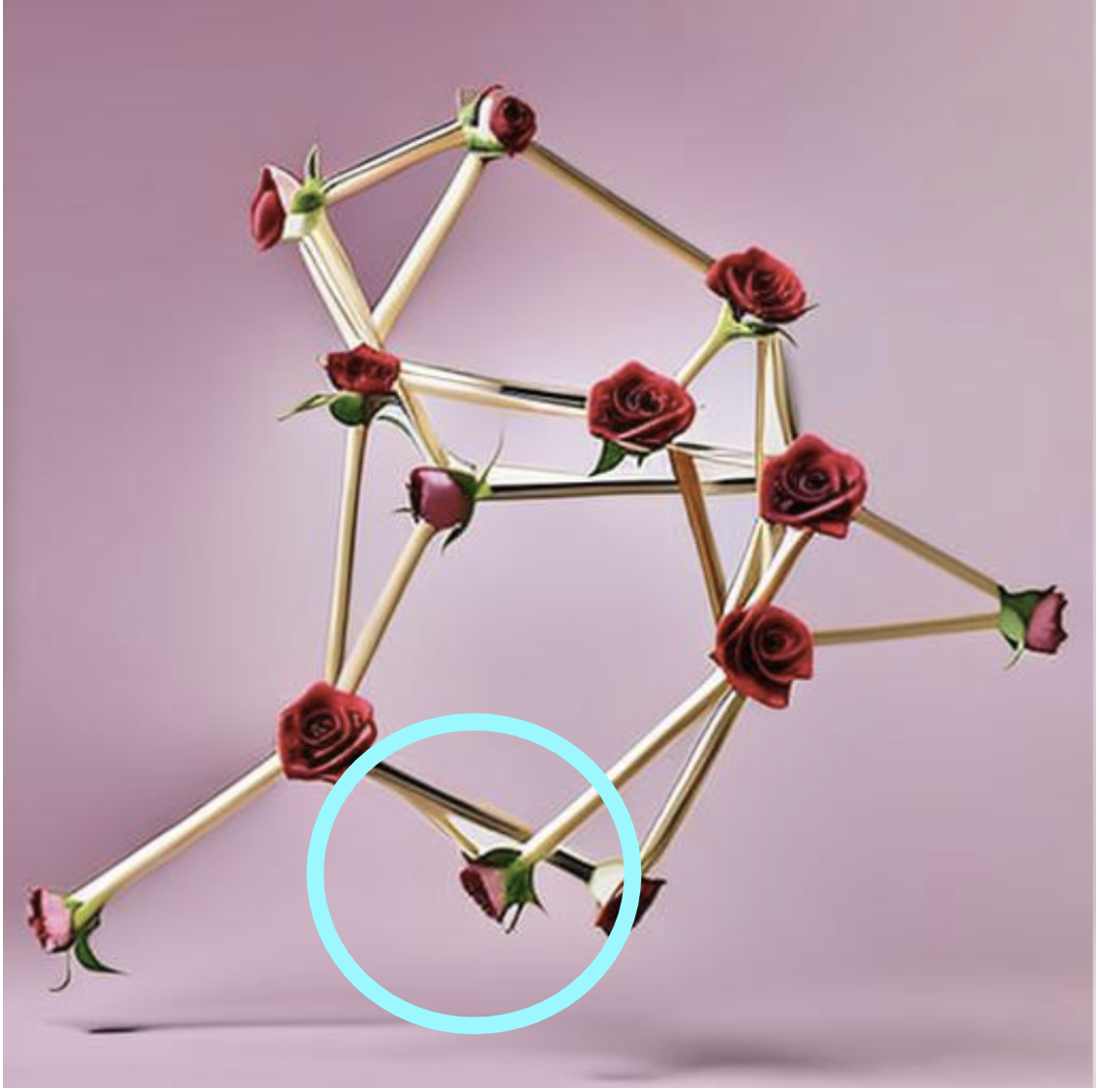}
         \caption{ \sys result using Depth2Img*}
         \label{fig:five over x9}
     \end{subfigure}

          \begin{subfigure}[b]{0.15\textwidth}
         \centering
         \includegraphics[width=1\textwidth]{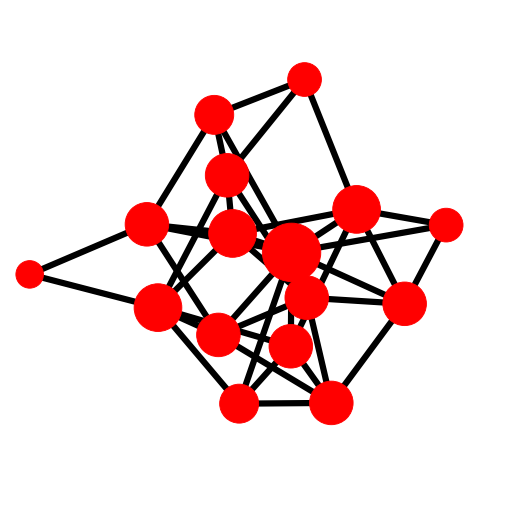}
         \caption{plain visualization}
         \label{fig:y equals x10}
     \end{subfigure}
     \hfill
     \begin{subfigure}[b]{0.15\textwidth}
         \centering
         \includegraphics[width=1\textwidth]{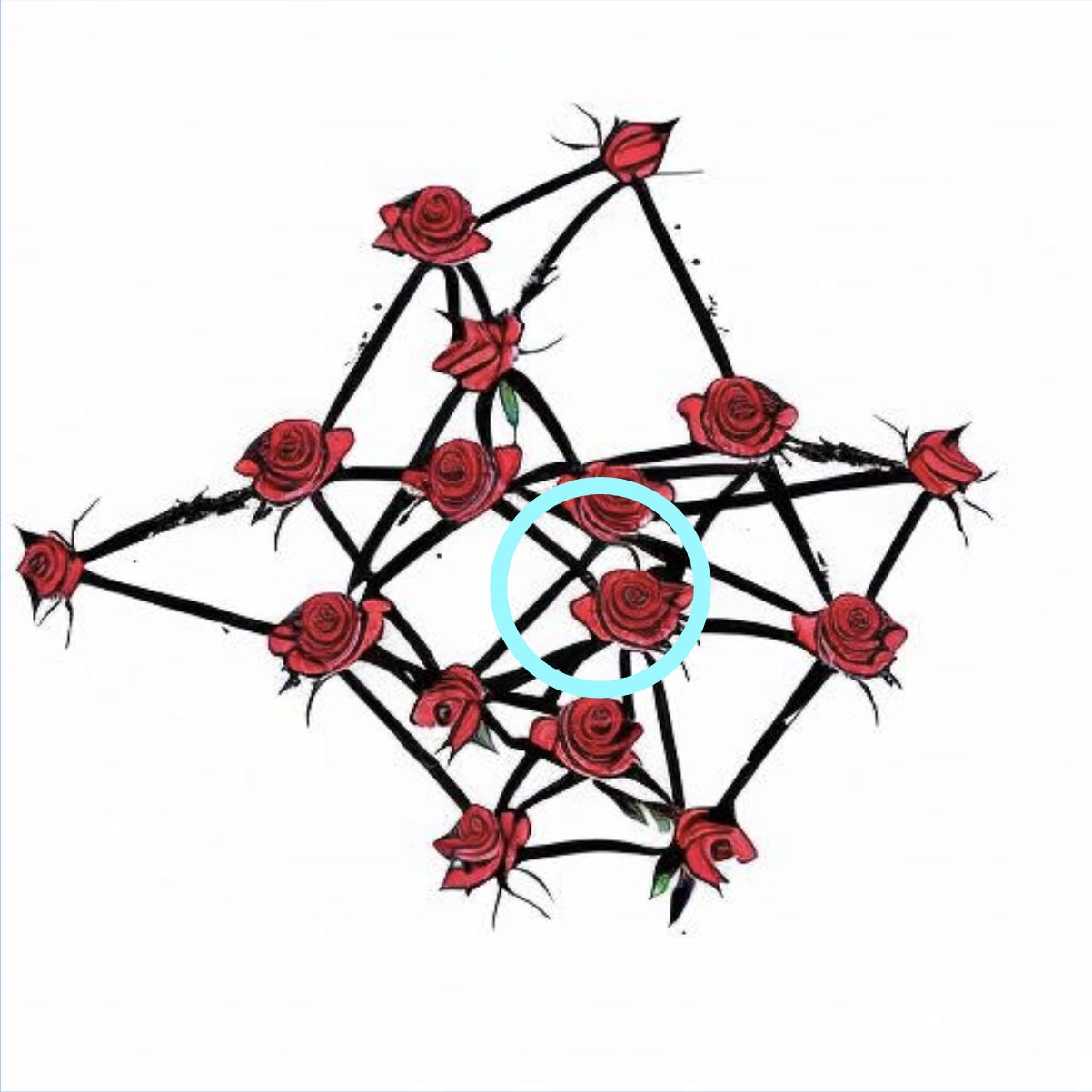}
         \caption{\sys result using Img2Img}
         \label{fig:three sin x10}
     \end{subfigure}
     \hfill
     \begin{subfigure}[b]{0.15\textwidth}
         \centering
         \includegraphics[width=1\textwidth]{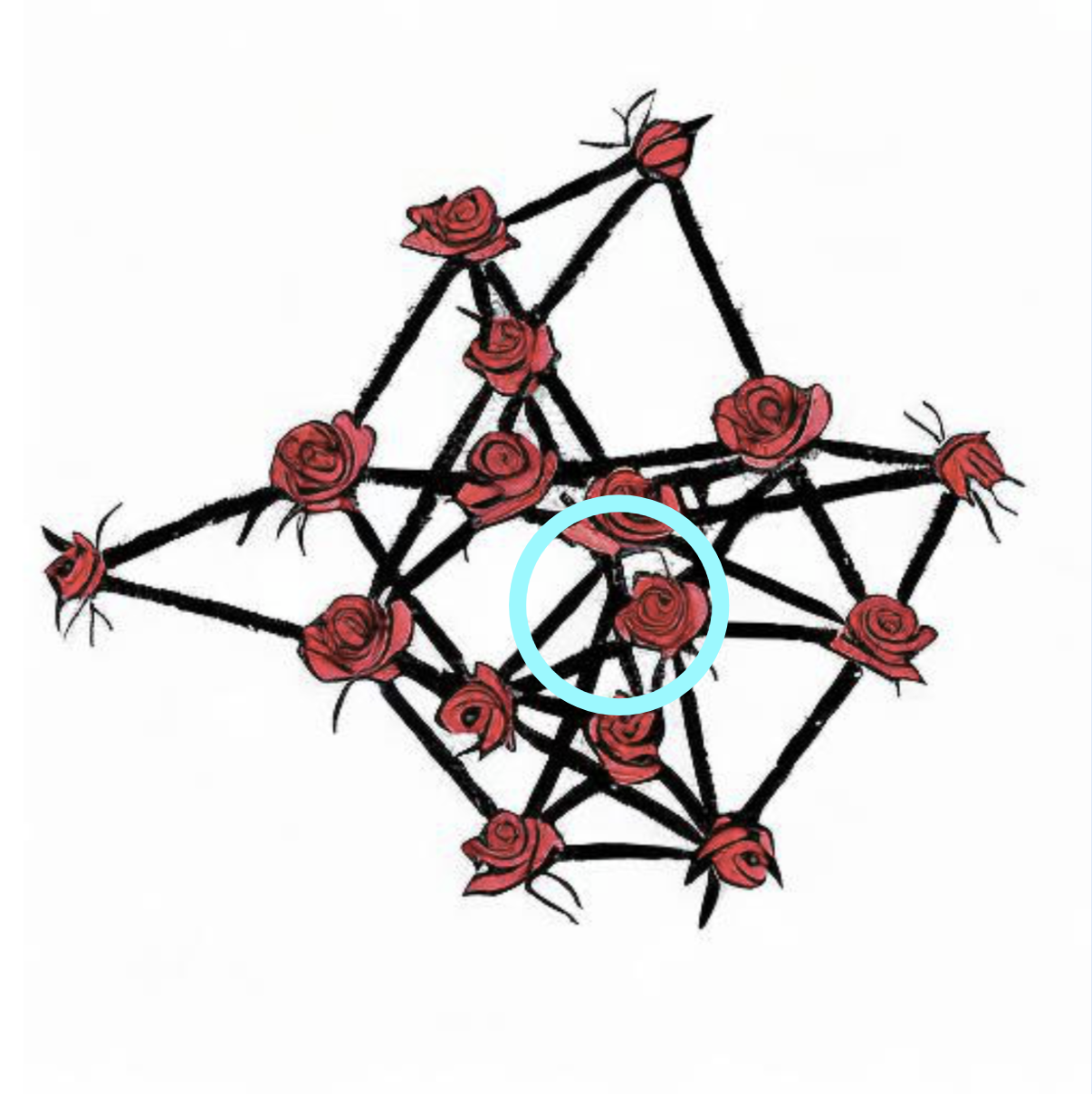}
         \caption{\sys result using Img2Img*}
         \label{fig:five over x10}
     \end{subfigure}
        \caption{Result comparison of original depth2img/img2img and depth2img*/img2img*. Prompt for (b) and (c): \textit{realistic red roses on a background of pink velvet}. Prompt for (e) and (f): \textit{hand drawn style roses}. The cyan circle highlights where edge information is enhanced.}
        \label{fig_edgeenhance}
\end{figure}

\paragraph{Edge-enhanced depth2img and img2img}
\label{Edge-enhanced Depth2Img and Img2Img}
The advantage of depth2img is that it preserves structural information robustly while combining stylized marks. With img2img, we have found that it can have high performance on non-realistic styles. Unfortunately, for network graphs, both pipelines tend to spoil `fragile' edges. To deal with this, we have a slightly augmented version of these pipeline (depth2img* and img2img*, respectively). These overlay the edges of the original network visualization on top of the initial smoothed sketch visualization. The result is that edges that vanished in the sketch are `forced' back into the image. This strategy can maintain edge information in the synthesis visualization (Figure~\ref{fig_edgeenhance}).

\subsubsection{The Refine step}
While visualizations generated in the Synthesize step can be an improvement over sketches (in that marks `interact'). However, the image can be low-quality with noisy spots. In these cases, we have found that an extra run of a low-strength depth2img pipeline with quality-improvement prompts (e.g., \textit{high resolution realistic clear photograph, 4k} yields a better final image. Note that these additional prompts are concatenated to the full prompt (i.e., the sub-prompts and background prompts). We used depth2img specifically, as the Refine step is usually done for visualizations that are intended to be more realistic. In other scenarios, the refine step is optional and can be replaced with any resolution-enhancing pipelines (e.g., super-resolution\footnote{\url{https://huggingface.co/docs/diffusers/api/pipelines/stable_diffusion/upscale}}).

\subsection{\sys Variations}
\label{Detailed Workflow Illustration}
Depending on the visualization type and prompt, different variants of \sys can be used (see Figure~\ref{fig_wholediagram}). We provide some intuition over why a specific workflow recipe may be selected (e.g., structure and shape of the marks, the idiom, and the realism of the desired output). We emphasize the unique properties of each workflow that requires modification to the \sys recipe. We emphasize that while each workflow is different, there are sub-elements that re-occur and may be useful for use in new idioms.

\subsubsection{Network Graph}
The \textbf{Sketch step} for network graphs uses the img2img diffusion pipeline on the $N \times N$ grid of the nodes. With multiple sub-prompts, \sys performs this step multiple times. For the \textbf{Synthesize step} we have found that network graphs require an initial img2img pipeline to smooth the sketched image.

\begin{figure}[t] 
     \centering
     \begin{subfigure}[b]{0.15\textwidth}
         \centering
         \includegraphics[width=1\textwidth]{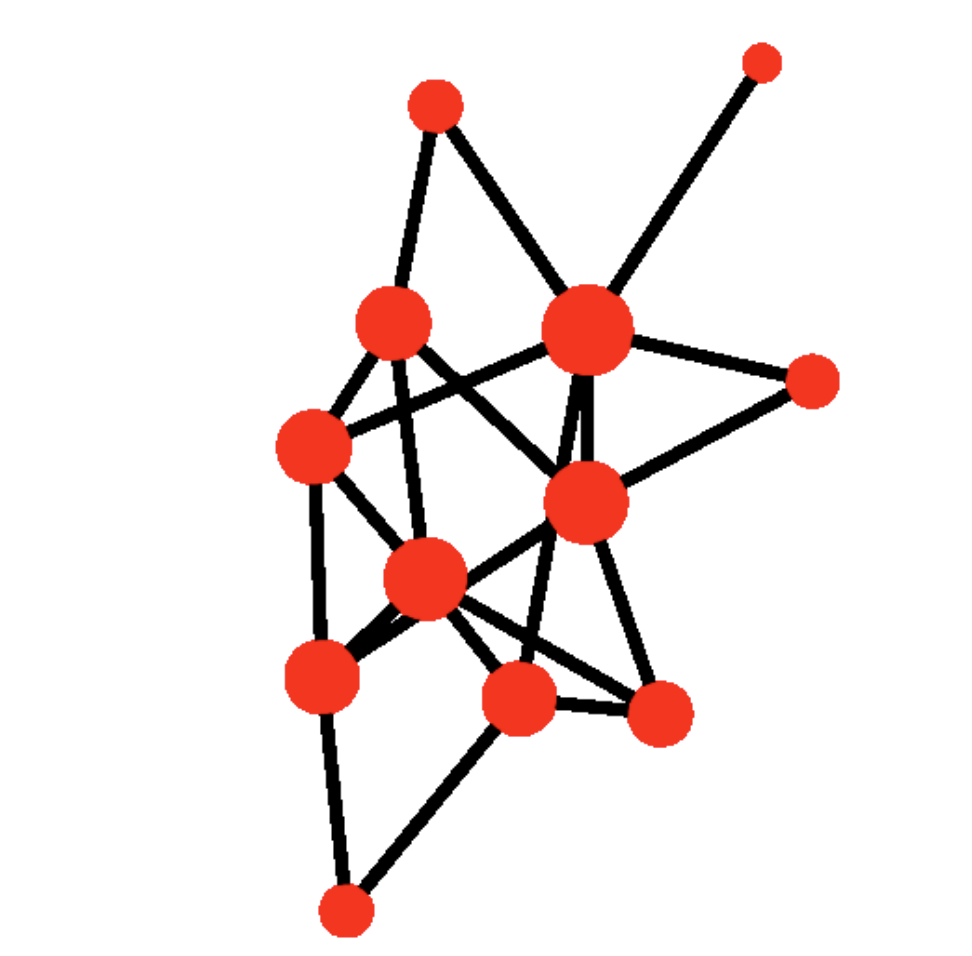}
         \caption{plain visualization}
         \label{fig:y equals x2}
     \end{subfigure}
     \hfill
     \begin{subfigure}[b]{0.15\textwidth}
         \centering
         \includegraphics[width=1\textwidth]{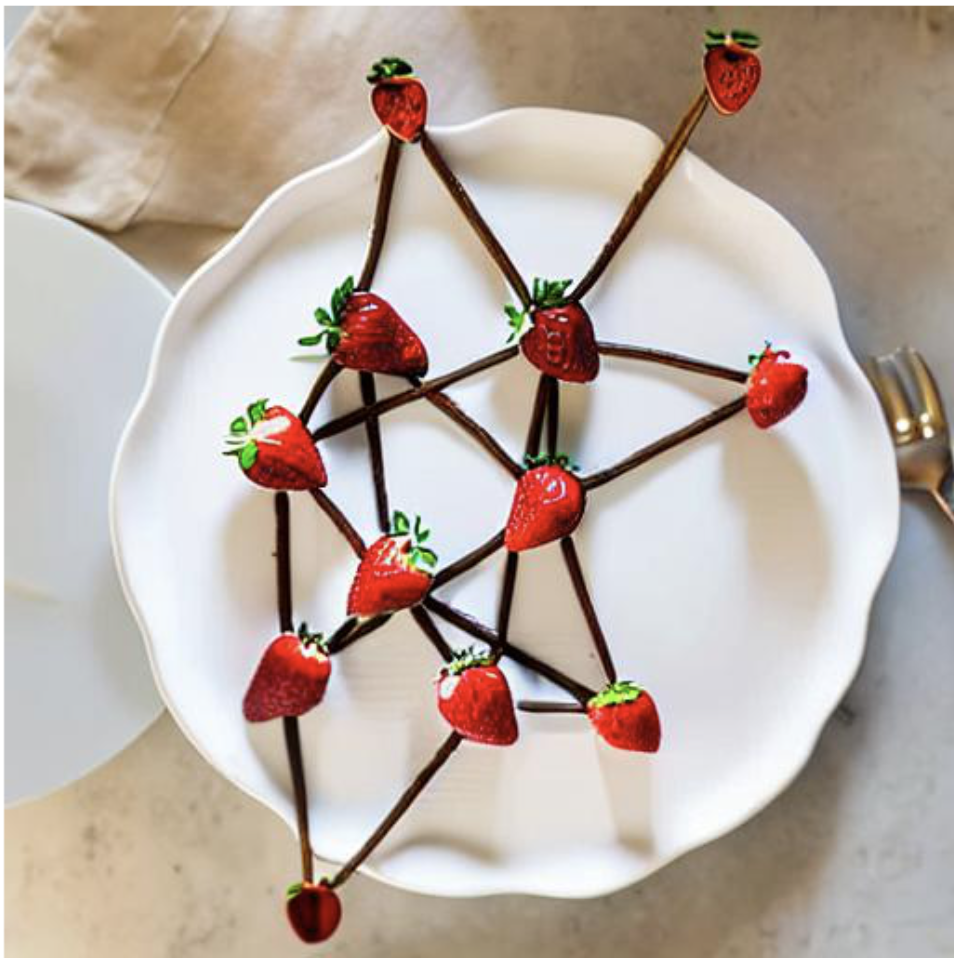}
         \caption{\sys result using $\text{ControlNet}^{\text{c}}$}
         \label{fig:three sin x1}
     \end{subfigure}
     \hfill
     \begin{subfigure}[b]{0.15\textwidth}
         \centering
         \includegraphics[width=1\textwidth]{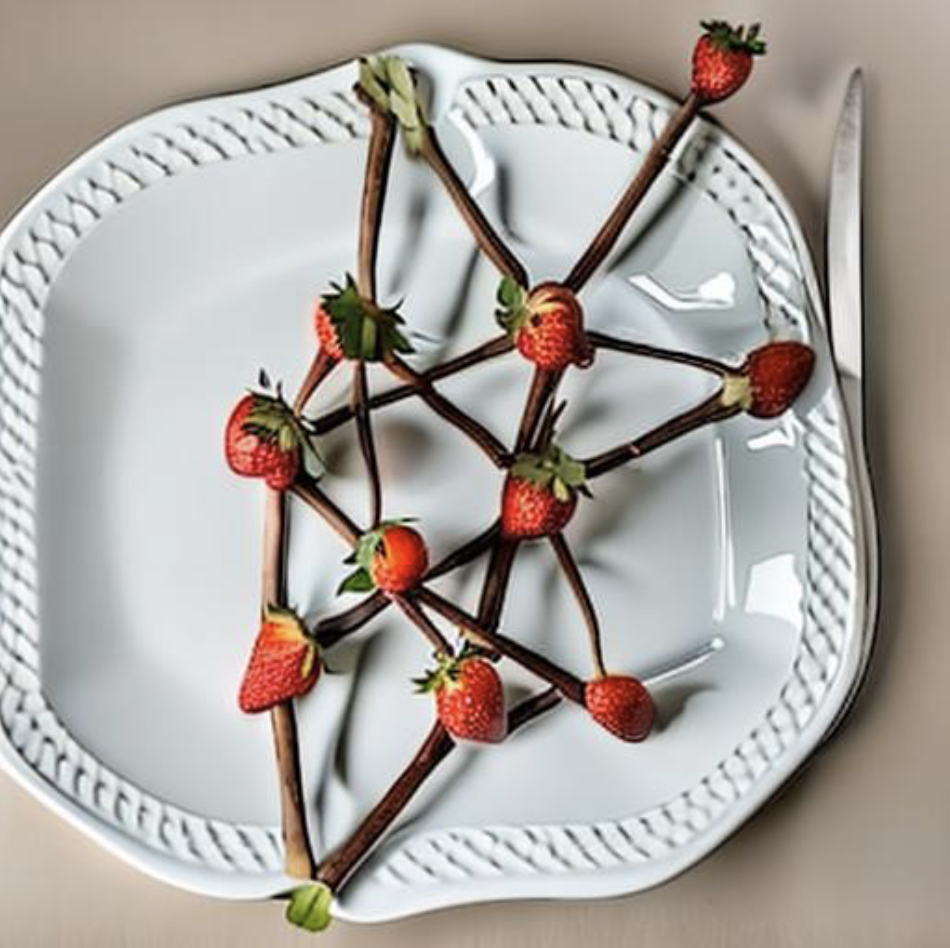}
         \caption{ \sys result using Depth2Img*}
         \label{fig:five over x2}
     \end{subfigure}

          \begin{subfigure}[b]{0.15\textwidth}
         \centering
         \includegraphics[width=1\textwidth]{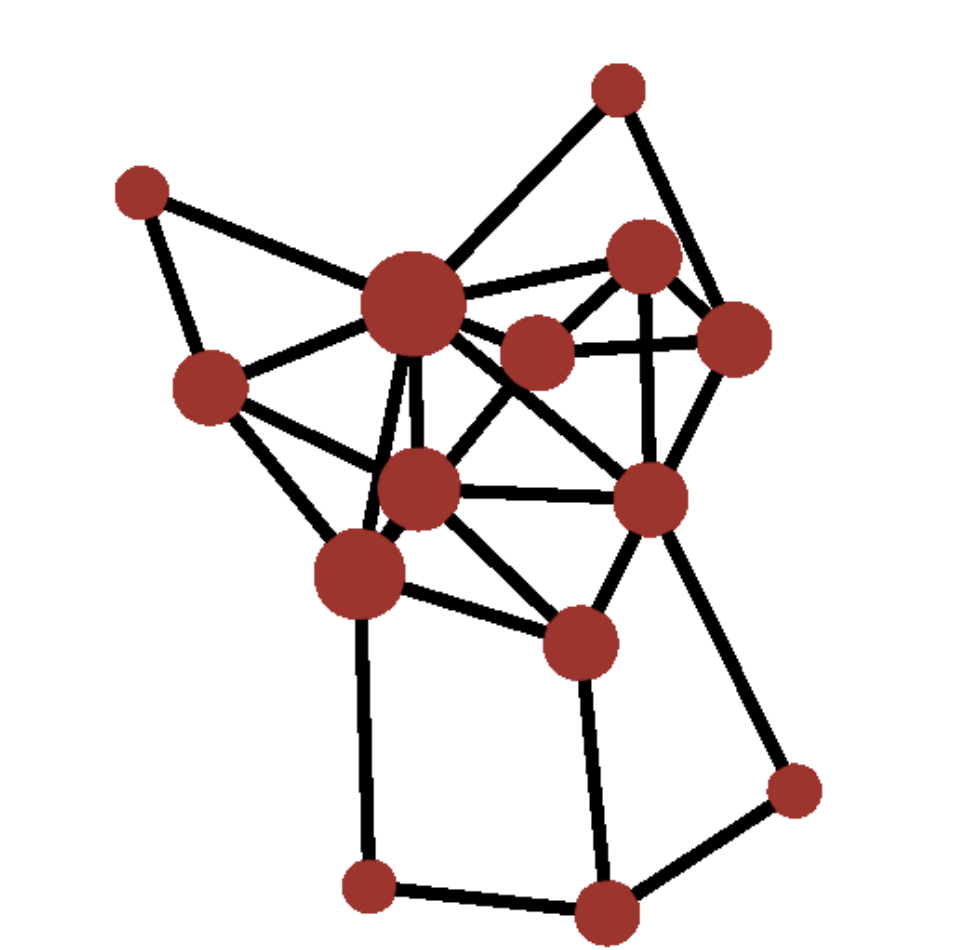}
         \caption{plain visualization}
         \label{fig:y equals x3}
     \end{subfigure}
     \hfill
     \begin{subfigure}[b]{0.15\textwidth}
         \centering
         \includegraphics[width=1\textwidth]{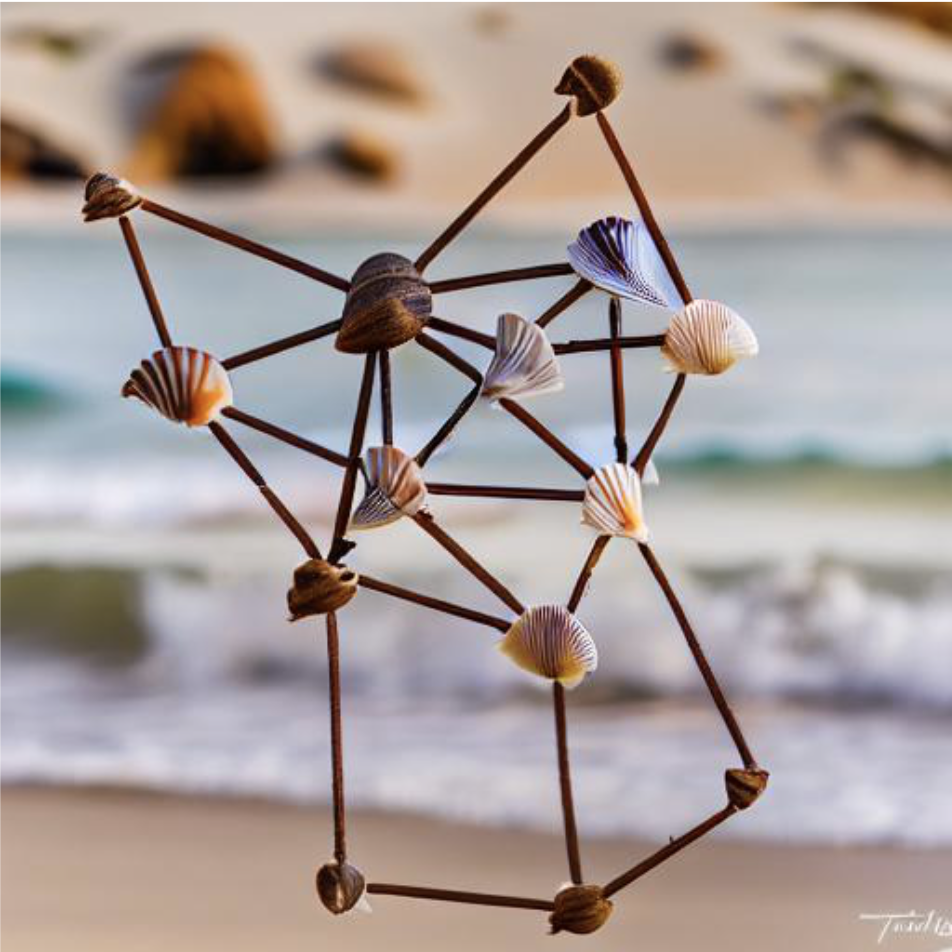}
         \caption{\sys result using $\text{ControlNet}^{\text{c}}$}
         \label{fig:three sin x2}
     \end{subfigure}
     \hfill
     \begin{subfigure}[b]{0.15\textwidth}
         \centering
         \includegraphics[width=1\textwidth]{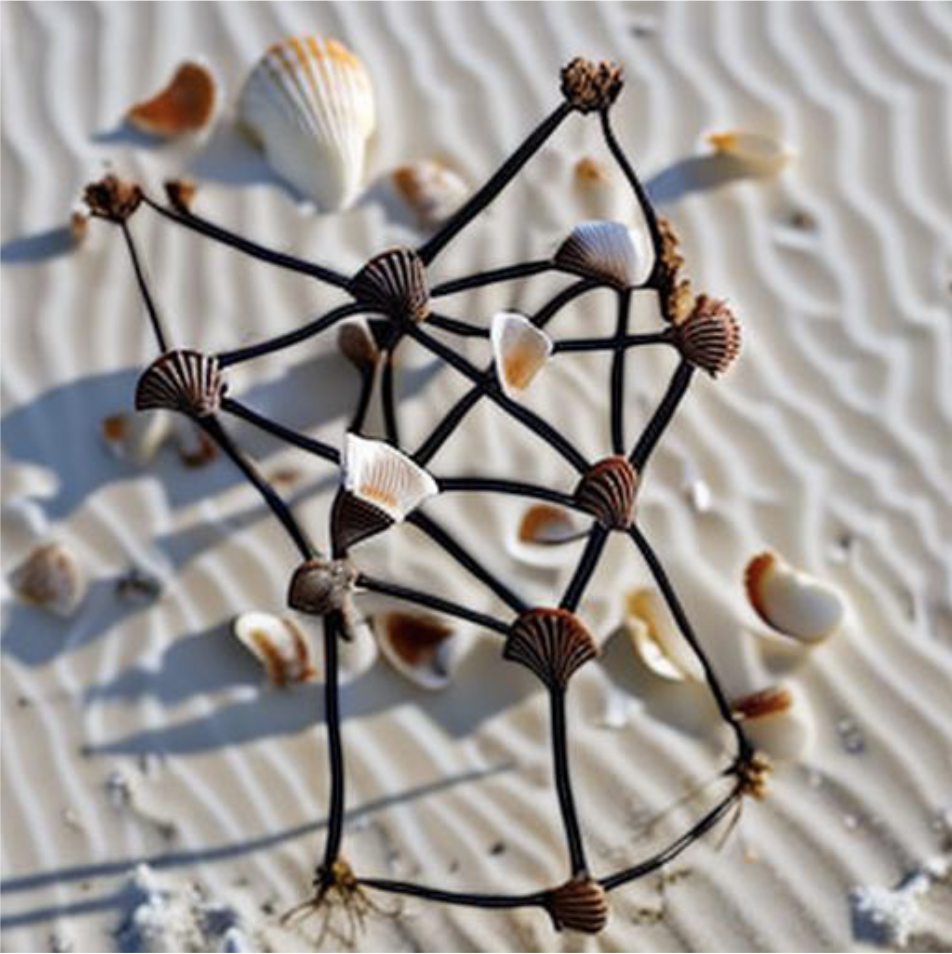}
         \caption{\sys result using Depth2Img*}
         \label{fig:five over x4}
     \end{subfigure}
        \caption{Result comparison of $\text{ControlNet}^{\text{c}}$ and depth2img*. Prompt for (b) and (c): \textit{realistic strawberries and branches on a white plate}. Prompt for (e) and (f) \textit{realistic seashells and branches on the white beach}}
        \label{fig_controlnetdepth2imgcompare}
\end{figure}

After this step, \sys has options depending on whether the prompt calls for a realistic or non-realistic output. For a realistic prompt, the diffusion pipeline can be $\text{ControlNet}^{\text{c}}$ (ControlNet with canny edge condition) or depth2img*. Using $\text{ControlNet}^{\text{c}}$ can result in more precise mark shapes, but depth2img* can produce a more stylized and visually cohesive result (See the comparison in Figure.~\ref{fig_controlnetdepth2imgcompare}). For non-realistic styles, the second diffusion pipeline can either be $\text{ControlNet}^{\text{c}}$ or img2img*. As with realistic images, $\text{ControlNet}^{\text{c}}$ can result in better visual cohesiveness, as well as more smooth edge structure. However, img2img can produces results that are more in line with prompt and have a more `sketched' look.   We have found that network graphs do not require an additional \textbf{Refine step}.

\subsubsection{Bar Chart}
Bar charts, which generally use single separated marks, work well with the following recipe. As with network graphs, in the \textbf{Sketch step}, an img2img pipeline is applied to a grid of equally-sized bars as a starting point. However, unlike the nodes in the network (which are generally equal-sized marks), bar charts require generated marks to be elongated or duplicated before being re-assembled. 

We omit smoothing in the \textbf{Synthesize step} as sketch visualizations for bar charts are generally clean. For realistic prompts, we have found that the DMP pipeline works best for bar charts. For non-realistic images, img2img produces better results. As DMP tends to produce some low-quality artifacts, we used a low-strength depth2img in the \textbf{Refine step} to fix these.

\subsubsection{Pie Chart}
Pie charts require a distinct Sketch step but otherwise can roughly follow the same workflow as the bar chart. This assumes that the pie segments can be cleanly separated.
Unlike bar charts and network nodes, pie slices are more complex in shape. Depending on the prompt, \sys can use depth2img (on the original visualization) or text2img (from scratch) in the \textbf{Sketch step}. In both cases, \sys will cut the pie slice from this generated content before re-assembly. The remaining steps, \textbf{Synthesize} and \textbf{Refine}, follow the same format as the bar chart.

\subsubsection{Area Chart}
Area charts are like pie charts in that the mark complexity is higher. We used the same \textbf{Sketch} and \textbf{Refine} steps as pie charts. For the \textbf{Synthesize} step we have found that both realistic and non-realistic images can be produced with either the DMP or $\text{ControlNet}^{\text{C}}$ pipeline. Because area marks tend to be large, this yields a simple depth structure that can work with $\text{ControlNet}^{\text{C}}$ and produce precise mark shapes. DMP can yield more stylized and visually cohesive results. Supporting both pipelines allows users to balance their trade-offs and produces a better stylized visualization.

\begin{figure}[t] 
     \centering
     \begin{subfigure}[t]{0.33\linewidth}
         \centering
         \includegraphics[width=1\textwidth]{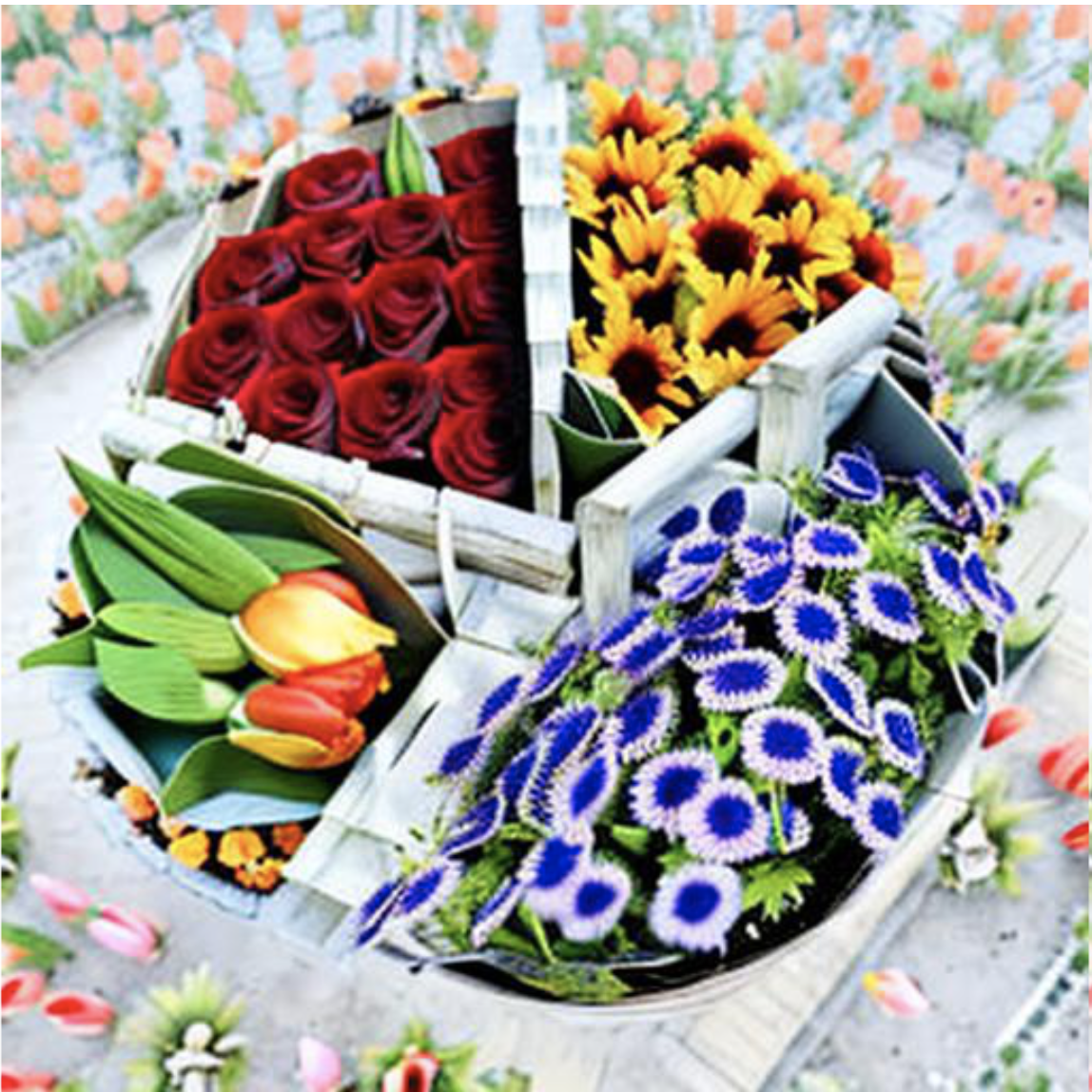}
         \caption{Refined visualization}
         \label{fig:y equals x1}
     \end{subfigure}
     \hfill
     \begin{subfigure}[t]{0.33\linewidth}
         \centering
         \includegraphics[width=1\textwidth]{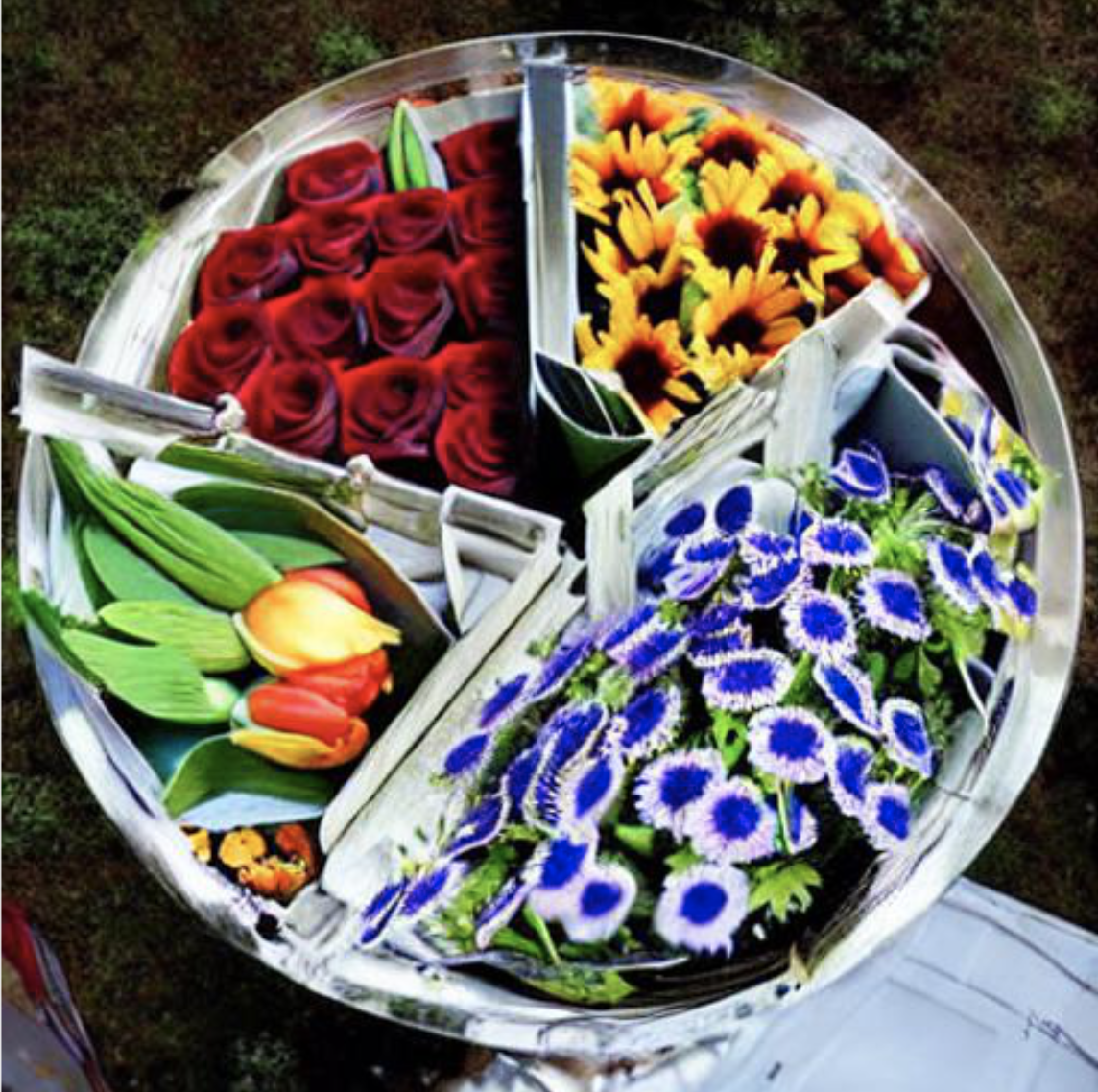}
         \caption{Re-generate background}
         \label{fig:three sin x}
     \end{subfigure}%
     \hfill
     \begin{subfigure}[t]{0.33\linewidth}
         \centering
         \includegraphics[width=1\textwidth]{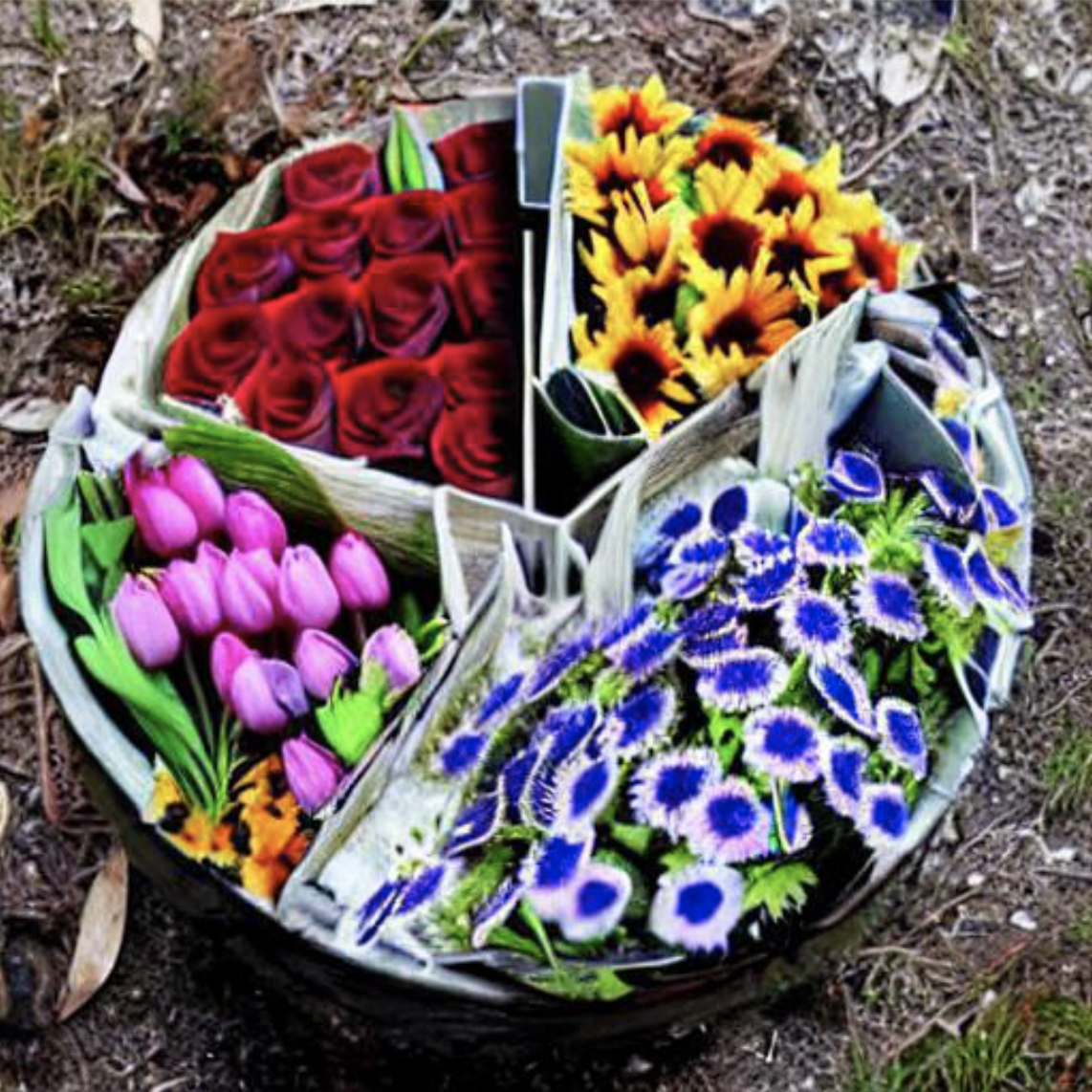}
         \caption{Re-generate background and a stylized slice}
         \label{fig:baseline2a}
     \end{subfigure}     

        \caption{Post-Processing by re-generating background and stylized mark(s). Used prompts are \textit{bird eye view of a field of realistic red roses/tulips/purple daisy flowers/orange sunflowers}.}
        \label{fig_postprocess}
\end{figure}

\subsubsection{Optional Post-Processing}
In some cases, \sys generates a visualization that generally has the stylization we want but may have failed for one sub-prompt (i.e., we like everything except the stylized hamburger bar). To fix these, a user can apply an optional inpainting pipeline to regenerate a specific mark or the background. When applied, \sys performs inpainting with the prompt and the mask corresponding to the selected mark/background. Examples of this step are in Figure~\ref{fig_postprocess}. Users can also choose to regenerate the visualization with a self-defined area and prompts that were not used in the initial generation. Finally, if parts of the visualization are well stylized, the user can copy those to worse areas and execute an inpainting pipeline with low-strength.
Clearly, additional edits may be needed or desired to produce a `finished' visualization (e.g., changing fonts, modifying the axes to match the new style, etc.).

\subsection{Implementation Details}
We implemented \sys with Python and Pytorch. We used Google's Colaboratory notebooks to specify \sys recipes. Plain visualizations are generated with Matplotlib, NetworkX and Altair. Image processing, such as mask generation, is achieved with OpenCV and Pillow. We used diffusion models with Diffusers from Hugging Face, using SD models. Specifically, the ControlNet pipeline uses SD v1.5\footnote{\url{https://huggingface.co/lllyasviel/ControlNet}}, and all other pipelines use SD v2\footnote{\url{https://huggingface.co/stabilityai/stable-diffusion-2-1}}. All images are 512$\times$512 in size. 

With internal experiments, we found some parameters that work well as defaults. Although \sys can work well with default parameters, these values can be fine-tuned depending on specific needs. We set guidance scale for all pipelines to 20. In the Sketch step, depth2img strength is set to 0.98 and img2img to 0.82. In the Synthesize step, img2img strength is set to 0.4, while all other pipelines (DMP, depth2img, ControlNet) used 0.8. In the Refine step, a value between 0.25 and 0.35 worked best as a default strength. For DMP pipeline, by default $\beta$ is set to 0.5.

\section{Evaluation}

To evaluate \sys, we considered three baselines (depth2img, img2img, and $\text{ControlNet}^{\text{c}}$) on four dimensions: 1) how well is the data preserved (information preservation)?; 2) how well the approach follows the prompt (prompt alignment)?; 3) how visually aesthetic is the generated visualization (aesthetic)?; and 4) how visually cohesive is the output (cohesiveness)? In all, we used three different prompts (both with single and multiple sub-prompts) for each of the four chart types (12 ``groups'' in total). We allowed each system to generate four possible outputs with the default parameters\footnote{For baselines, default strengths are set to 0.8 (img2img) and 0.95 (depth2img, $\text{ControlNet}^{\text{c}}$), as these values showed the best overall performance.}. 
One author created the samples and randomized the results (but did not participate in the evaluation).
For all conditions, we used the negative prompt of ``low quality, normal quality, worst quality, poorly drawn, error, abstract, blurry.''
For \sys, the workflow that best fits with the visualization type and style specified in the prompt was selected (i.e., realistic or non-realistic). 
When there are two possible workflows, the higher quality workflow was chosen (which would be what the users of \sys would likely do). 
Note that with \sys, we did not perform the optional post-processing.
Our prompts came from different sources. Some were descriptions of existing stylized visualizations, and others were suggested by designers from the Data Visualization Society or our research groups. Three evaluators (two male, one female) who were blind to the generation conditions evaluated each of the 192 generated images on the four metrics using a five-level Likert scale. 
For the analysis of each metric, we performed the Kruskal-Wallis test with all conditions, followed by Dunn's Test as a post hoc analysis.

\begin{figure}[t] 
     \centering\includegraphics[width=\linewidth]{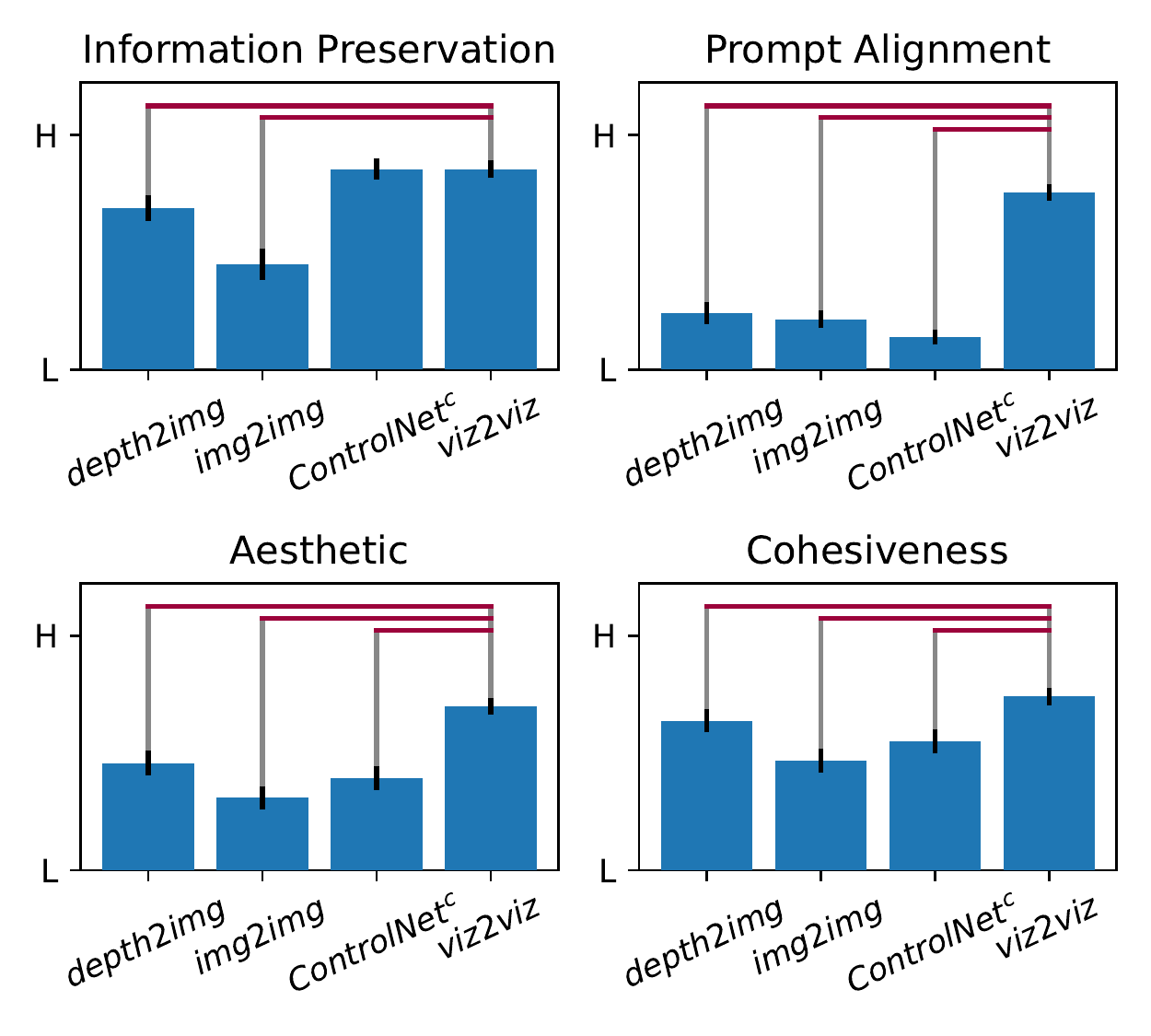}
    \caption{The evaluation results. The error bars indicate 95\% confidence intervals, and the red connecting lines show which conditions are significantly different from \sys.}.  
    \label{fig_result}
\end{figure}

Our results are summarized in Figure~\ref{fig_result}. 
For `preservation of information,' we saw a significant difference ($H=115.7$, $p<0.001$), with \sys significantly outperforming depth2img ($p<0.001$) and img2img ($p<0.001$). Naturally, ControlNet follows the outlines of the original visualization well (which is why we use it as a subcomponent of \sys). With `prompt alignment' ($H=270.8$, $p<0.001$) \sys has significantly higher quality than all other conditions ($p<0.001$ for all). 
The trend was similar for the aesthetics metric ($H=117.6$, $p<0.001$) with \sys scoring significantly higher than all other conditions ($p<0.001$ for all).
Finally, for cohesiveness, conditions showed a significant difference ($H=60.9$, $p<0.001$), with \sys again outperforming all others ($p<0.05$ for depth2img, and $p<0.001$ for others).
The average Spearman's $\rho$ value for each metric was 0.20, 0.31, 0.23, and 0.13 for information preservation, prompt alignment, aesthetic, and cohesiveness, respectively. These indicate a slight to fair agreement between the annotators.

As a case study, we took the images we generated using DALL-E 2 (see Figure~\ref{fig:taxonexamples}) and tried to generate close approximations using \sys. In most situations, we replicated a close approximation with \textit{real} data as input. We describe examples that were less successful below. Prompts and materials from both studies are include in the supplementary materials.

\section{Discussion and Future Work}

\subsection{Generalizability of \sys}
We designed \sys to be generalizable to other visualization idioms. The three general recipe steps (Sketch-Synthesize-Refine) are intended to balance the preservation of visualized information while coherently styling the visualization. As our experience shows, we believe that there is an inherent challenge in creating one pipeline that works across all inputs and prompts. That said, we believe that in addition to the high-level workflow's generalizability, sub-components in our specific implementations can be applied to new visualization types. For example, a scatter plot can be built on the network graph's workflow and a donut chart can be based on the pie chart.

\subsection{Prompt Engineering}
Visualizations in \sys are highly sensitive to the quality of the prompts. While we leave a systemized evaluation of prompt engineering to future work, we consider some patterns we have observed.

In \sys, prompts work better on more specific and common objects than on abstract or imaginary objects. For example, \textit{a side view of a wooden roller coaster} is more likely to get a better result over \textit{side view 'of cosmic red roller coaster, surreal, psychedelic themes}. Things that the underlying diffusion model is bad at generating are unlikely to work for stylized visualizations. 
For example, we have found that models struggle with \textit{medical syringe}.
Testing the prompt without any data points in a standard pipeline (e.g., text2img, DALL-E, etc.) may be useful in tuning prompts that have a better chance of working. For example, we note that \textit{fries} tends to have better results than \textit{chips} (as this meaning is somewhat more rare). In some cases, a single prompt can have ambiguity in what it will render, and additional text is useful. For example, adding \textit{fruit} or \textit{color} to \textit{orange} will improve the chances of getting the desired stylization. Finally, we find that there are differences in capabilities of different generative algorithms. For example, Stable Diffusion struggled with a branded Coca-Cola bottle.

As with generative prompts, in general, we find that including detailed descriptions of \textit{the camera view, the number of objects, styles, colors, the specific type of objects, shapes, object names, layouts, and background} can help. 
Explicitly describing the desired style or medium (e.g., \textit{photorealistic} or \textit{oil painting}) appears to help. When working with one-to-many style visualizations, adding \textit{knolling} can be useful. One area in which we have not extensively experimented is with negative prompts. We use these in the final `cleanup' steps of \sys, but it is possible that such prompts can help guide the system to produce better results in earlier operations.

\subsection{Getting High Quality Images}
Our experience with \sys (and other generative systems) is that getting a good result often takes multiple attempts. In this paper, we have attempted to capture key decision points that would lead one to choose a particular workflow or pipeline over another. We have also tried to acknowledge the various scenarios in which human intervention or repetition is necessary. We have found in our experiments that the upper bound on finding a good image (given a good prompt) is around 10 images (though many take far less).

Not all prompts can be easily generated. When color or shape of the plain-visualization mark is so different from the desired object, the generation might fail. Thus, some coherence is required between initial inputs and desired outputs (e.g., a red pie chart slice will work much better for the prompt \textit{a slice of pizza} over a yellow slice).

In addition to prompts, most pipelines have other tunable parameters. We have offered good defaults based on our experiments. However, it is likely that certain stylization requests would benefit from further tuning. For example, a user may want a good value for the $\beta$ parameter in DMP (0.5 by default). In response to intermediate results (e.g., the quality of the Sketched visualization), the user can increase (if the image is promising) or decrease this value. Finally, we emphasize that \sys can fail completely or partially. Using post-generation inpainting and similar features can help recover from partial failures. %

\subsection{Limitations and Future Work}

Currently, \sys is implemented in the context of a notebook environment. While this can be easily used as a library, an important area to tackle in the future is the implementation of a GUI. Being able to easily see the effect of changes, seeing intermediate results, varying parameter spaces, and generally having access to direct manipulation operations (e.g., for inpainting) will likely lead to better outputs.

While \sys can generate many different types of marks, sometimes better semantic understanding (both of language and objects) can lead to better outcomes. For example, it is difficult to specify that we would like the bars in a chart to be represented by the amount of liquid in a syringe (rather than the syringe's length). One potential approach is to use sketched or found examples as additional input. Because we do not tightly control the interaction between foreground and background objects, \sys is unable to generate certain infomages~\cite{coelho2020infomages} (the top-hat in Figure~\ref{fig:taxonexamples}, for example). Future work, with additional steps to pre-generate the background and specify the area in which the visualization should be rendered, may allow \sys to support these stylized visualizations more broadly.

As with other generative systems, some pipelines can produce noisy results. With \sys the problem can become more pressing as our various generation constraints can lead to unfortunate side-effects in what the visualization looks like. Future work to address this may involve better parameter tuning, pipeline creation, better masks (for DMP and inpainting) and improved interfaces.

Finally, \sys is currently limited to the scale of marks it can support. As the number of marks increases, they may begin to blend improperly during diffusion. There is nothing inherently problematic in generating smaller subsets of the visualization in pieces. Characterizing the limits of this approach and ensuring, technically, that images can be reconstructed in a way that looks good would be areas of future work.

\section{Conclusions}
In this paper, we introduce \sys, a generation workflow allowing users to get high quality stylized visualizations given a text prompt and plain visualization. We introduce a three-step recipe that ensures that the stylized visualization encodes data correctly while at the same time satisfying complex criteria like image cohesiveness. To achieve this, we make use of state-of-art controllable diffusion models as well as our self-designed pipeline (DMP). We describe how the \sys can be generalized to other stylization tasks. We believe that there is both an inherent challenge and opportunity in using generative approaches in visualization. We are excited to offer a possible entry point into this area for both practitioners and researchers. 

\acknowledgments{We would like to thank Jason Forrest, Elsie Lee-Robbins, and Matt Brehmer, as well as early pilot users who gave us prompts and feedback. This work was partially funded by NSF IIS-1815760.}

\balance 
\bibliographystyle{abbrv-doi-hyperref-narrow}

\bibliography{reference}

\pagebreak
\appendix
\begin{figure*}[htp]
 \centering 
\includegraphics[width=0.85\textwidth]{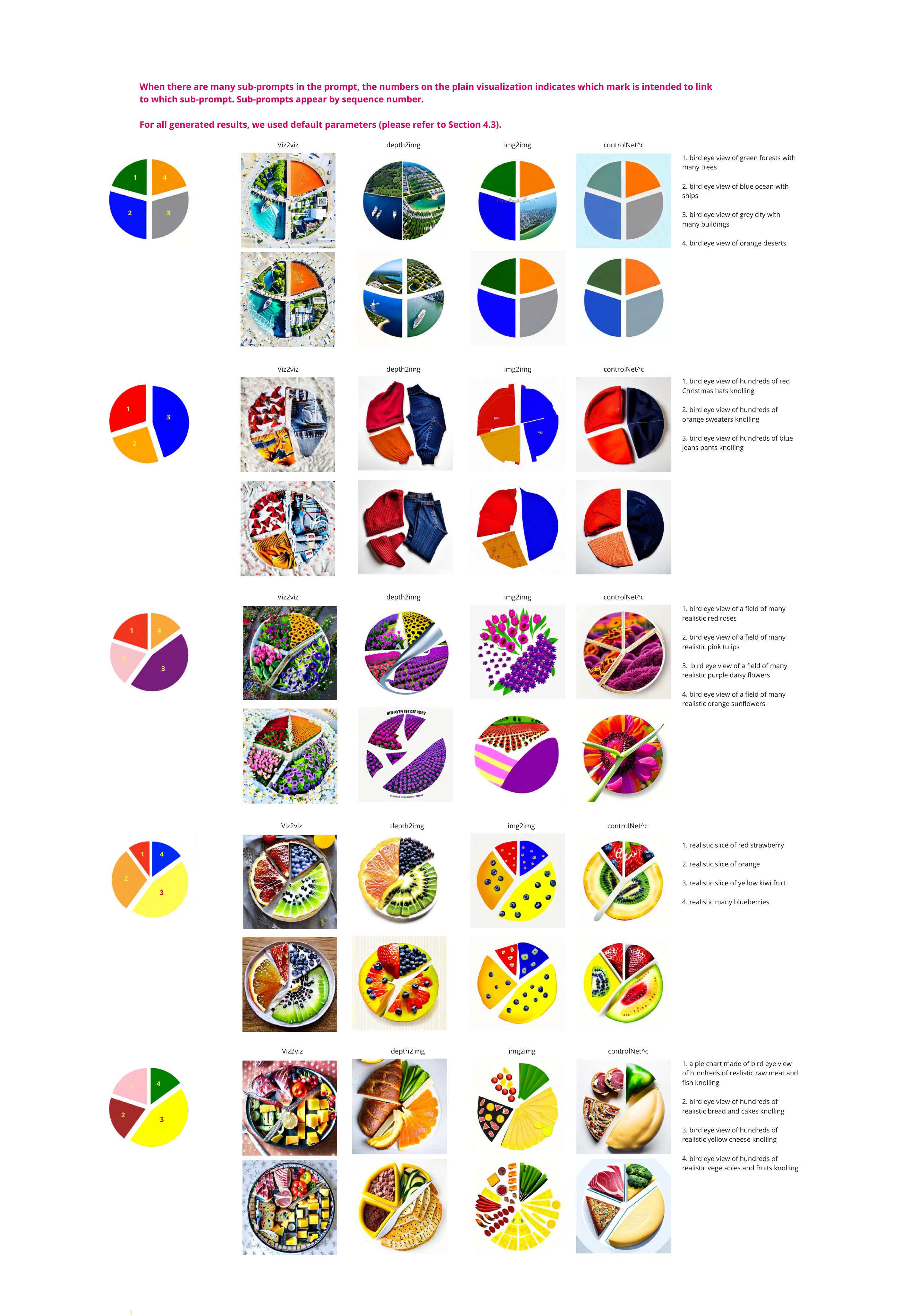}
\caption{Supplementary results 1.}
\end{figure*}

\pagebreak
\begin{figure*}[htp]
 \centering 
\includegraphics[width=0.85\textwidth]{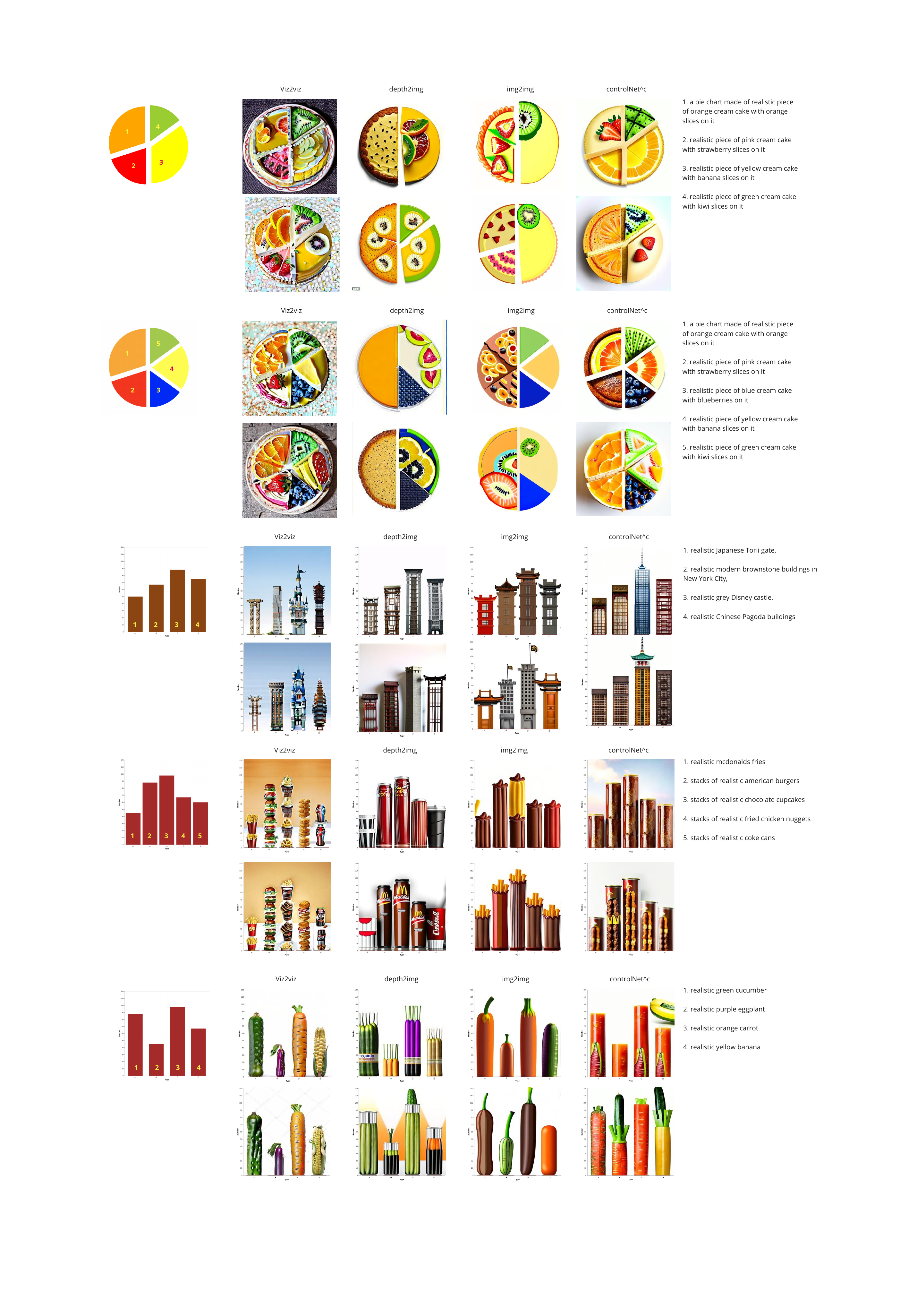}
\caption{Supplementary results 2.}
\end{figure*}

\pagebreak
\begin{figure*}[htp]
 \centering 
\includegraphics[width=0.85\textwidth]{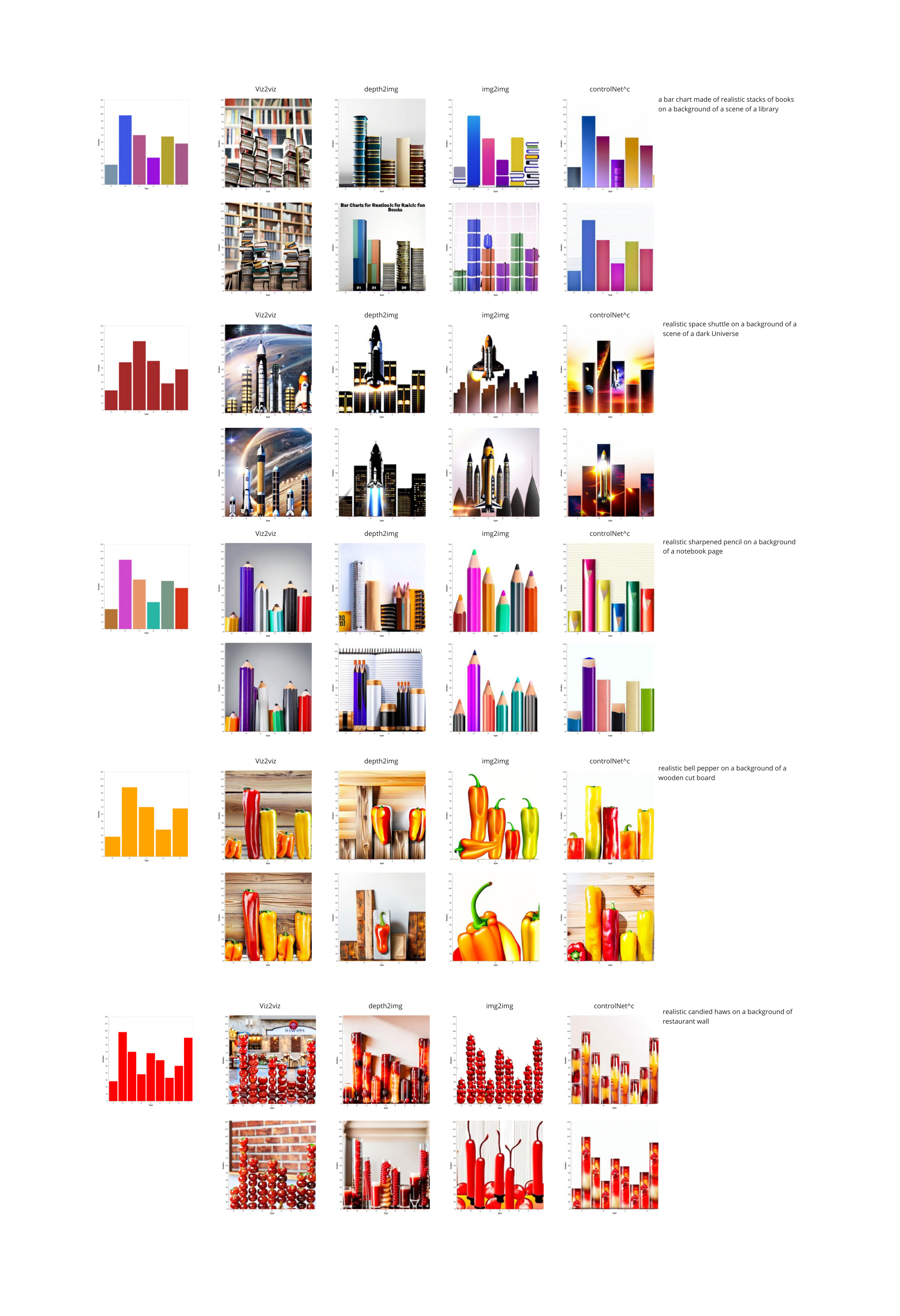}
\caption{Supplementary results 3.}
\end{figure*}

\pagebreak
\begin{figure*}[htp]
 \centering 
\includegraphics[width=0.85\textwidth]{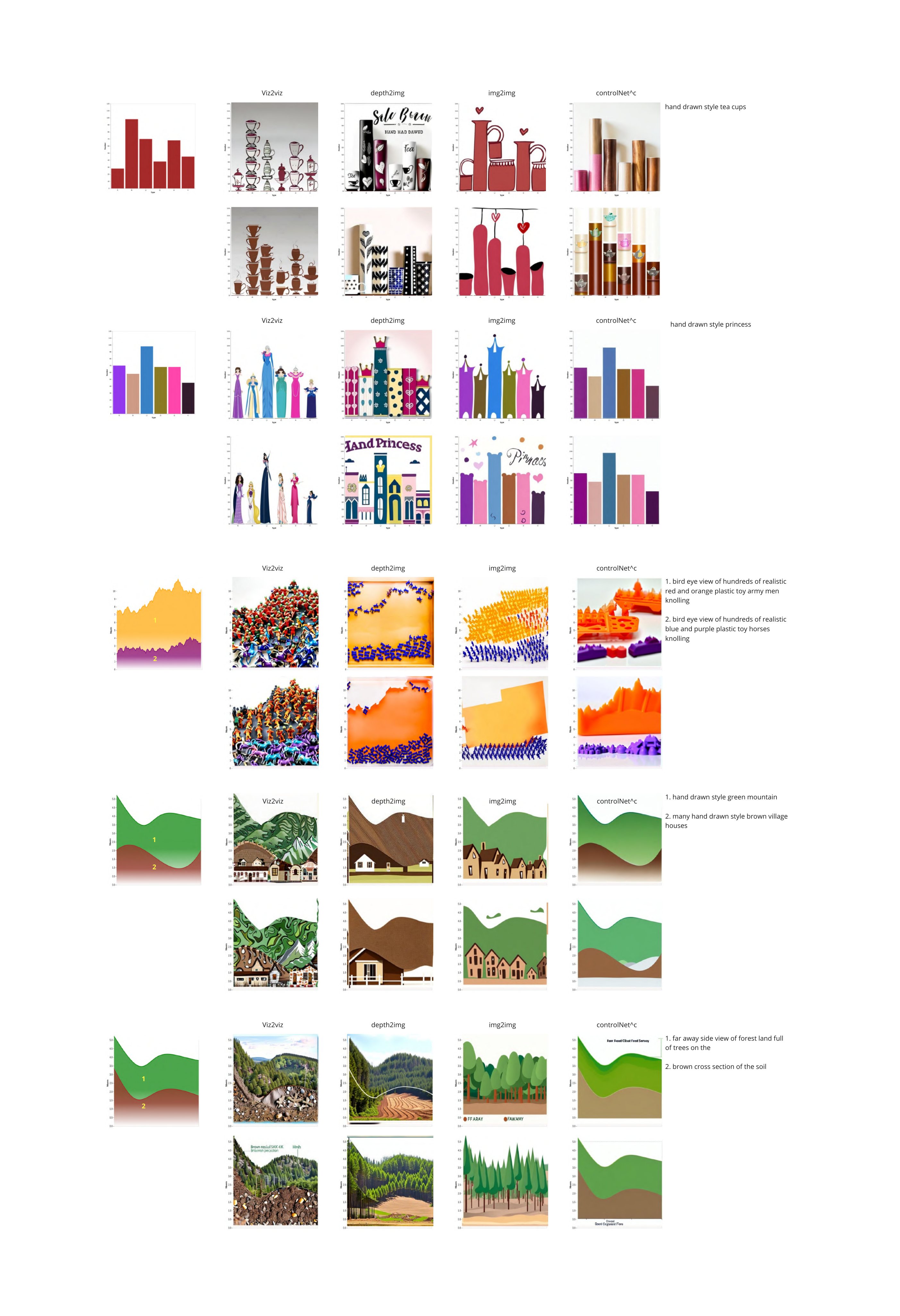}
\caption{Supplementary results 4.}
\end{figure*}

\pagebreak
\begin{figure*}[htp]
 \centering 
\includegraphics[width=0.85\textwidth]{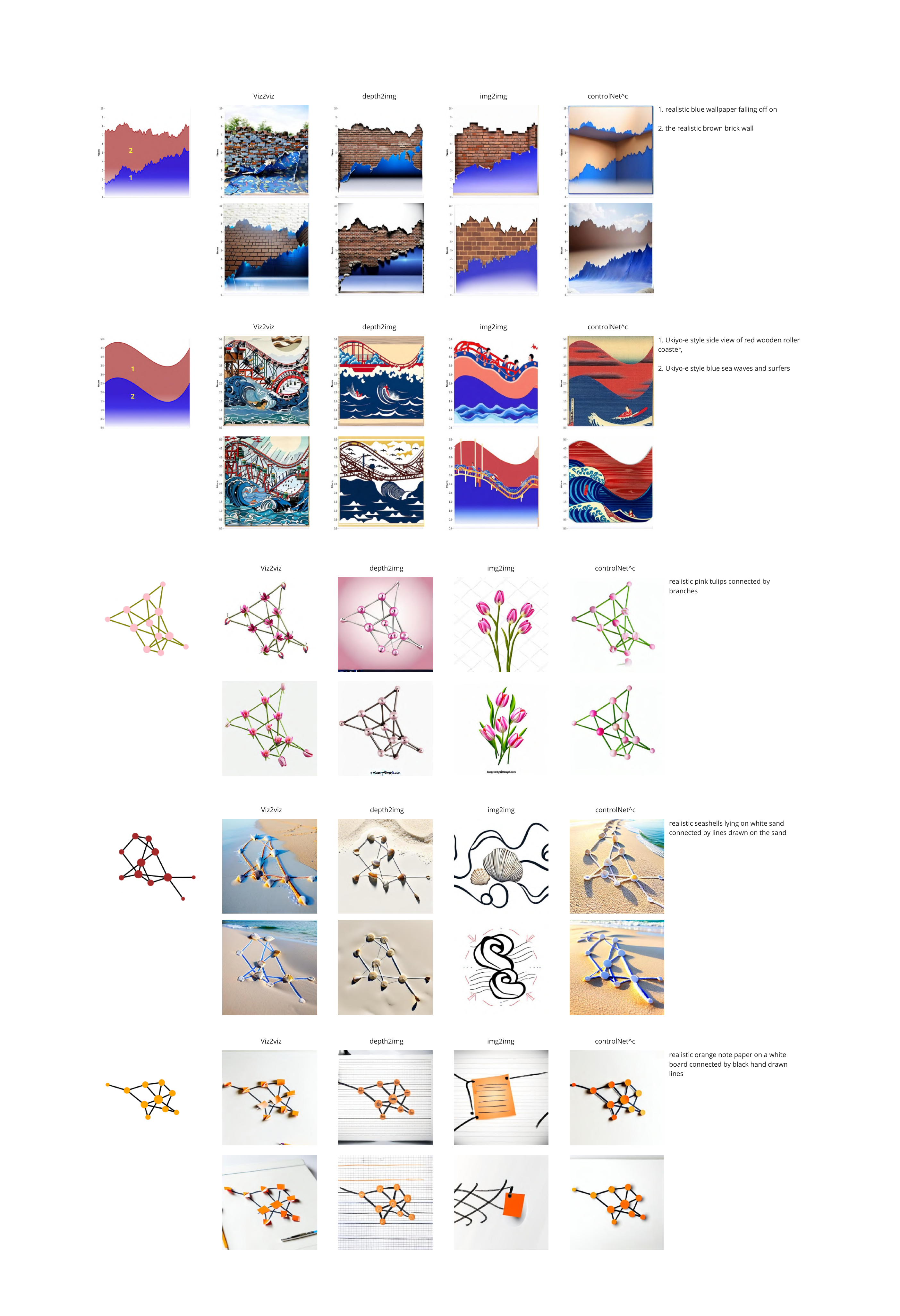}
\caption{Supplementary results 5.}
\end{figure*}

\pagebreak
\begin{figure*}[htp]
 \centering 
\includegraphics[width=0.85\textwidth]{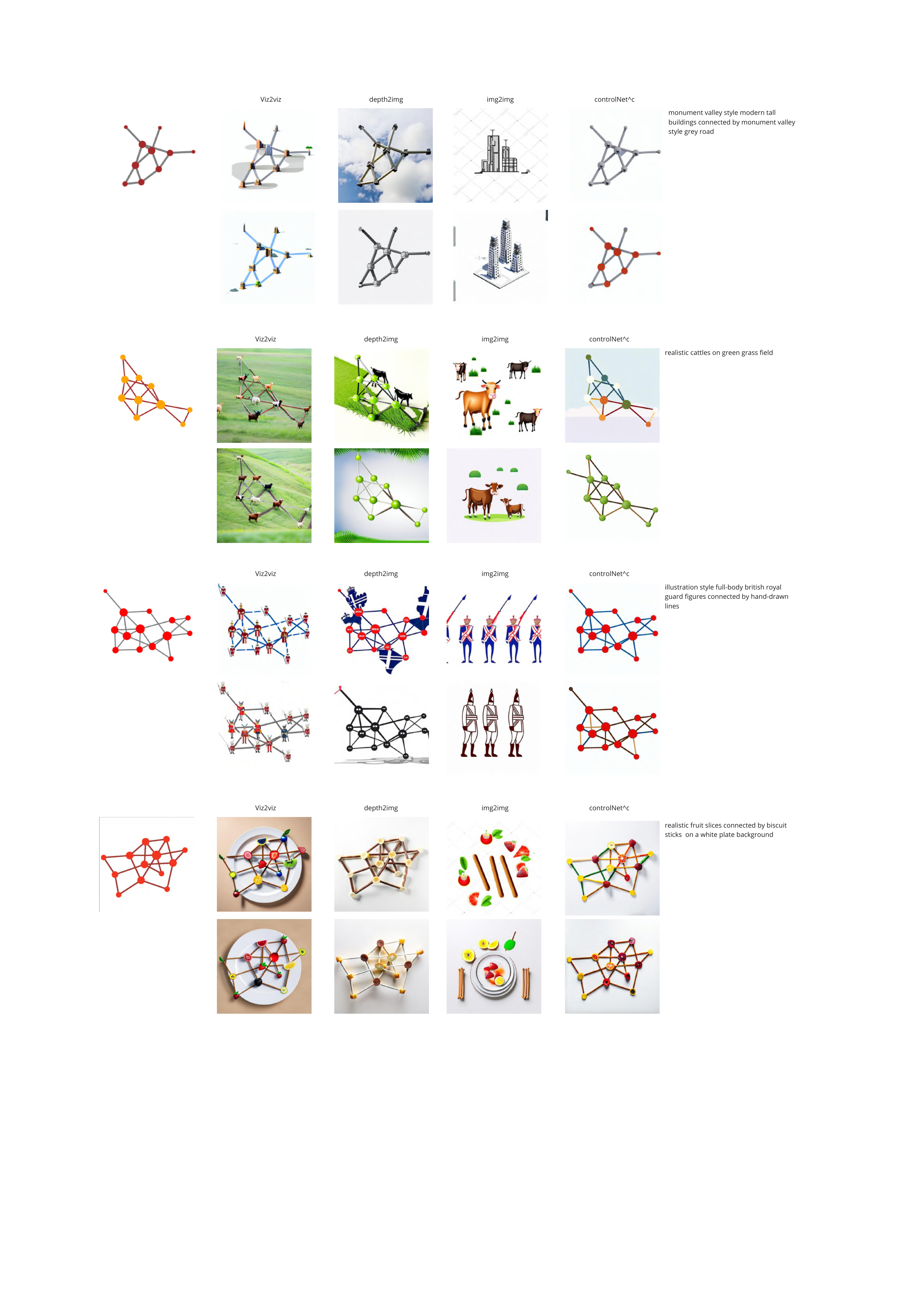}
\caption{Supplementary results 6.}
\end{figure*}

\pagebreak
\begin{figure*}[htp]
 \centering 
\includegraphics[width=0.85\textwidth]{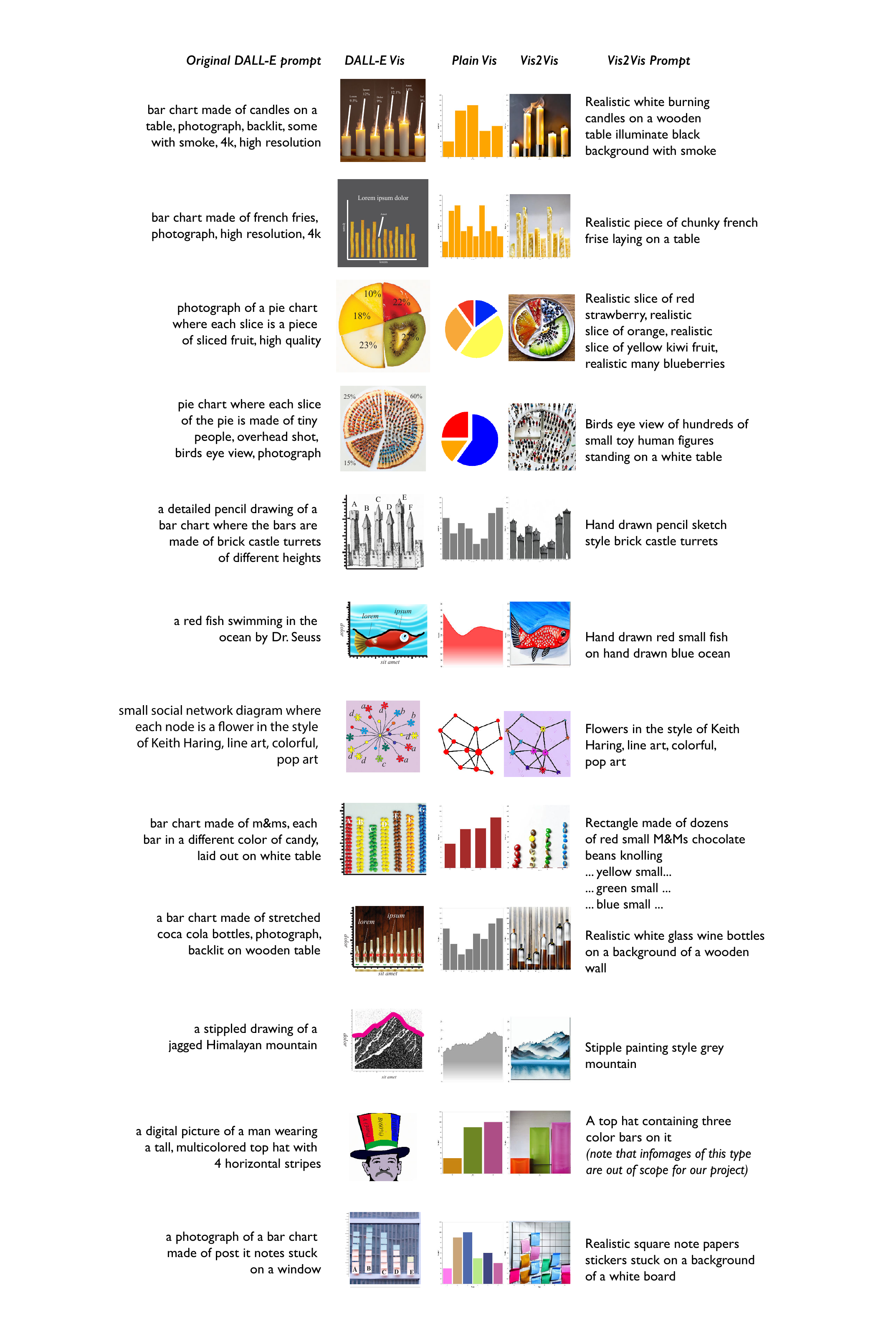}
\caption{Comparison between DALL-E 2 and \sys. Note that DALL-E 2 cannot specify the underlying data while \sys can.}
\end{figure*}

\end{document}